\documentclass[12pt,a4paper,english,superscriptaddress,aps,nofootinbib]{revtex4}
\usepackage[utf8]{inputenc}
\usepackage[T1]{fontenc}
\usepackage{amsmath,amssymb,graphicx}
\makeatletter
\usepackage{babel}
\usepackage[active]{srcltx}
\usepackage{graphicx,color}
\usepackage{changebar}
\usepackage{hyperref}
\usepackage[T1]{fontenc}
\usepackage{esint}
\usepackage{multirow}
\usepackage{dsfont}
\usepackage{ae}
\usepackage{amsmath}
\usepackage{braket}
\usepackage{mathtools}
\usepackage{slashed}
\usepackage{empheq}
\usepackage{multirow}
\usepackage{graphicx}
\usepackage{amsfonts}
\usepackage{amsmath}
\begin{document}
\title{Regular black holes in Einstein Cubic Gravity}

\author{L. A. Lessa}
\email{E-mail: leandrolessa@fisica.ufc.br}
\affiliation{Universidade Federal do Cear\'{a} (UFC), Departamento de F\'{i}sica - Campus do Pici, Fortaleza, CE, C. P. 6030, 60455-760, Brazil.}

\author{J. E. G. Silva}
\email{E-mail: euclides.silva@ufca.edu.br}
\affiliation{Universidade Federal do Cariri(UFCA), Av. Tenente Raimundo Rocha, \\ Cidade Universit\'{a}ria, Juazeiro do Norte, Cear\'{a}, CEP 63048-080, Brazil}

%\author{R. V. Maluf}
%\email{E-mail: r.v.maluf@fisica.ufc.br}
%\affiliation{Universidade Federal do Cear\'{a} (UFC), Departamento de F\'{i}sica - Campus do Pici, Fortaleza, CE, C. P. 6030, 60455-760, Brazil.}

%%%%%%%%%%%%%%%%%%%%%%%%%%%%%%%%%%%%
%\author{C. A. S. Almeida}
%\email{E-mail: carlos@fisica.ufc.br}
%\affiliation{Universidade Federal do Cear\'{a} (UFC), Departamento de F\'{i}sica - Campus do Pici, Fortaleza, CE, C. P. 6030, 60455-760, Brazil.}

%\author[ufca]{}
%\ead{r.v.maluf@fisica.ufc.br}
%%%%%%%%%%%%%%%%%%%%%%%%%%%%%%%%%%%%

%%%%%%%%%%%%%%%%%%%%%%%%%%%%%%%%%%%%
%\author[pici]{C. A. S. Almeida}
%\ead{carlos@fisica.ufc.br}

%\address[pici]{Universidade Federal do Cear\'a (UFC), Departamento de F\'isica, Campus do Pici, Fortaleza - CE, C.P. 6030, 60455-760 - Brazil}

%\address[ufca]{Universidade Federal do Cariri(UFCA), Av. Tenente Raimundo Rocha, \\ Cidade Universit\'{a}ria, Juazeiro do Norte, Cear\'{a}, CEP 63048-080, Brazil}

%%%%%%%%%%%%%%%%%%%%%%%%%%%%%%%%%%%%%

%%%%%%%%%%%%%%%%%%%%%%%%%%%%%%%%%%%%%

\begin{abstract}
%We construct the four-dimensional magnetically charged black hole in  Einstein Cubic Gravity (ECG).
We investigate the effects of the Einstein cubic gravity (ECG) on regular black hole solutions driven by nonlinear electrodynamics (NLE) sources.
The ECG tends to form a naked singularity at the origin for a high ECG coupling constant. Assuming that ECG provides only perturbative corrections to the regular magnetic charged solutions, we found modified regular solutions with a de Sitter-like core whose cosmological constant depends on the magnetic charge and the ECG coupling constant. 
%At the perturbative level, the black hole horizon is modified by the magnetic charge and thus, the thermodynamic functions as the Hawking temperature, heat capacity are modified by the ECG terms. 
The thermodynamic stability is investigated by means of the Hawking temperature and the heat capacity.
In fact, for a small charge and ECG coupling, the Hawking temperature is regularized, leaving a thermodynamic stable remnant for a small $r_h \neq 0$. The heat capacity reveals that the ECG regular black hole undergoes a phase transition between an unstable into a stable configuration.
%, increasing until it reaches a maximum and then, the temperature decreases until it vanishes for $r_h \neq 0$.
\end{abstract}

%\begin{keyword}
%Eisnteinian Cubic Gravity, Regular Black Hole, Nonlinear Electrodynamics
%\PACS 11.30.Cp \sep 11.15.-q \sep 11.30.Qc 
%\end{keyword}
\maketitle

%\linenumbers

\section{Introduction}

Black holes are one of the most striking predictions of the general relativity (GR). After the observation of gravitational waves (GW) from binary black hole mergers \cite{bhgw} and the data collected from the Event Horizon Telescope (EHT) \cite{eht}, new windows were open to probe Einstein gravity (EG) and its extensions in the strong field regime. Several modified gravity theories possesses black hole solutions, among them we highlight the $f(R)$ \cite{fr}, non-commutative \cite{nc} and Lorentz violating theories \cite{kr,ovgun}. These modified solutions may shed some light into viable quantum gravity models  \cite{kiefer}.

Among the expected features of modified gravity models, higher derivative theories naturally appears due to quantum corrections of the EG \cite{ren}.
In addition, string theory predicts the presence of higher curvature terms steaming from an effective low energy limit of the Born-Infeld term \cite{myers}. A particular higher curvature modified theory is the so called Lovelock theory constructed with a Lagrangian quadratic in the Riemann tensor whose equation of the motion (EOM) are of second-order \cite{lovelock}. In four dimensions, the Lovelock gravity is given by EG plus the Gauss-Bonnet term, and thus, the theory has the same degrees of freedom (DOF) of the EG. Modified Einstein-Gauss-Bonnet gravity theories with the same DOF of EG can lead to non-singular strong field solutions \cite{gb1,rgb,gb2}.

By adding cubic powers of the Riemann tensor, only one particular combination yields a new Lagrangian term $\mathcal{P}$, with the same DOF of the EG, at least at the weak field regime \cite{pablosp}. 
Accordingly, this higher-order theory was named \textit{Einstein cubic gravity} (ECG) \cite{pablosp}. In the strong field regime, the ECG still produces singular black holes for the static solutions \cite{pablos, robie}. Moreover, stationary \cite{spining1,spining2} and charged modified black hole solutions \cite{extremal,charged,charged2,charged3,spining3} were also found in ECG. In cosmology, ECG dynamics can produce both an early and late expansion epochs \cite{cosmology1,cosmology2,cosmology3,cosmology4}, and with a bounce between two De Sitter vaccua \cite{bouncecosmology}. The ECG effects on quasinormal modes (QNM) \cite{qnm}, shadows \cite{shadow}
and gravitational lensing were also discussed \cite{lensing}.

The resolution of spacetime singularities is another motivation for studying modified gravitational theories. Although supported by the well-known singularity theorems \cite{s1,s2}, the singularities can be avoided in EG by adding non-standard sources. Indeed, Sakharov \cite{r1} and Gliner \cite{r2} proposed that a vacuum configuration leading to a De Sitter core near the origin could lead to a regular black hole \cite{dym}. Bardeen \cite{bardeen} proposed a regular black  hole solution which latter Ayon-Beato-Garcia \cite{ag,ag2} and Bronikov \cite{bronikov1} showed to be the result of a nonlinear electrodynamics (NLE) as a source. 
Hayward constructed a regular black hole by assuming a particular \textit{mass function} \cite{hayward}. Fan and Wang constructed a family of regular black holes solutions in NLE which contains the Bardeen, Hayward and a new class solutions \cite{fan}. The interpretation of the parameters and the regimes of the solutions was latter improved by Bronikov \cite{bronikov} and Toshmatov et al \cite{bobir}. Regular black hole solutions were also found in modified gravity theories, such as in $f(R)$ \cite{regularmodified}, Lovelock \cite{regularlovelock}, Palatini \cite{regularpalatini}, Born-Infeld \cite{regulardbi} and GUP-corrected gravity \cite{maluf}. Although plagued with internal instabilities \cite{instability2}, regular black holes can be understood as effective smoothed quantum gravity corrected solutions near the origin \cite{frolov}. 

In this work we study how the ECG gravity modifies the regular black hole solutions. Assuming a two parameter dependent nonlinear electrodynamics Lagrangian which provides the magnetically charged Bardeen, Hayward and new class solutions, we numerically found the metric solution for the modified gravitational equations. It turns out that the ECG-NLE solutions exhibit a singular behaviour near the origin. For large values of the ECG coupling constant, even naked singularities are formed. Considering perturbative ECG effects, a balance between the ECG and NLE yields to regular black holes with a De Sitter core and a modified horizon. The thermodynamic stability of these solutions were analysed by means of the Wald entropy \cite{w1,w2} and the corresponding heat capacity. For small black holes, the temperature decreases until it vanishes for a non zero mass. As a result, stable remnant configurations was found. Likewise the EG regular solutions \cite{pablos,men,fan2,milko}, Lovelock \cite{my} and $f(T)$ \cite{miao} black holes, the entropy does not satisfy the usual Berkestein-Hawking expression, allowing thermodynamic stable states.

The work is organized as the follows. In the section \ref{sec2} 
we present the main features of the ECG, as well as the NLE Lagrangian we used as the source for the regular solutions. In Sec.\ref{s3} we obtain the static and spherically symmetric solutions of ECG-NLE dynamics, analysing the presence of singularities and horizons. 
In sec. \ref{sec4}, we investigate the perturbative effects of ECG on the regular Bardeen, Hayward and the new class solutions. The thermodynamic analysis and stability is performed in the Sec. \ref{sec5}. Final comments and discussion are outlined in the sec. \ref{sec6}. A number of useful results are summarized in the appendices. We work in $(3+1)$ dimension and with units where $c=1$.
%%%%%%%%%%%%%%%%%%%%%%%%%%%%%%%%%%%%%%%%%%%%%%%%%%%%%%%%%%%%%%%%%%%%%%%%%%%%%%%%%%%%%%%%%%%%%%%%%%%%%%%%%%%%%%%%%%%%%%%%%%%%%%%%%%%%%%%%%%%%%%%%%%%%%%%%%%%%%%%%%%%%%%%%%%%%%%%%%%%%%%%%%%%%%%%%%%%%%%%%%%%%%%%%%%%%%%%%%%%%%%%%%%%%

\section{Einstein cubic gravity coupled to a nonlinear electrodynamics}
\label{sec2}

We start by defining the action for the model we are studing.
Let us consider the Einstein cubic gravity (ECG) coupled to a nonlinear electrodynamics field (NED) of the type \cite{pablosp}
\begin{equation} 
\label{action}
   S = \frac{1}{16 \pi G} \int d^{4}x \sqrt{-g}\bigg[R- G^{2}\lambda \mathcal{P} + \mathcal{L} (F)  \bigg],
\end{equation}
where $G$ is the Newton constant and $\lambda$ is a dimensionless coupling constant which we will assume to be positive, $\lambda \geq 0$. The ECG term $\mathcal{P}$ is an invariant construct with contractions of the curvature tensor up to third order \cite{pablosp}. There is an unique 
combination of this cubic gravity which shares the same graviton spectrum with Einstein gravity (EG) in the weak field regime, and that has dimension-independent coupling constants, namely \cite{pablosp}
\begin{align} 
\label{p}
  \mathcal{P} = 12 R_{\mu}\ ^{\rho}\ _{\nu}\ ^{\sigma} R_{\rho}\ ^{\gamma}\  _{\sigma}\ ^{\delta} R_{\gamma}\ ^{\mu}\ _{\delta}\ ^{\nu} + R_{\mu \nu}^{\rho \sigma}R_{\rho \sigma}^{\gamma \sigma} R_{\gamma \sigma}^{\mu \nu} - 12 R_{\mu\nu\rho\lambda}R^{\mu\rho}R^{\nu\sigma} + 8 R_{\mu}^{\nu}R_{\nu}^{\rho}R_{\rho}^{\mu}.
\end{align}
The cubic gravity theory governed by the action in Eq.\ref{action} with $\mathcal{P}$ given by Eq.\ref{p} is called Einstein cubic gravity (ECG) \cite{pablosp}.

On the other hand, the $\mathcal{L}(F)$ term is a nonlinear electromagnetic field, where $F$ is the field strength of the vector field $A$ defined by $F := F_{\mu\nu}F^{\mu\nu} $, where $F_{\mu\nu}=\partial_{\mu}A_{\nu}-\partial_{\nu}A_{\mu}$. The most general ansantz for $A$ is given by
\begin{equation}
    A = \phi(r)dt + Q_m cos\theta d\phi,
\end{equation}
where $\phi(r)$ is a like-scalar potential of Maxwell and the $Q_m$ is the total magnetic charge defined by
\begin{equation}\label{total}
    Q_m = \frac{1}{4 \pi} \int F.
\end{equation}
Moreover, the equation of motion like-Maxwell and the Bianchi identities for NED, respectively, are given by
\begin{equation}
    D_{\mu}(\mathcal{L}_{F}F^{\mu\nu})=0  \  \  \    D_{\mu}(^{*} F^{\mu\nu})=0
\end{equation}
where $\mathcal{L}_{F}= \frac{\partial\mathcal{L}}{\partial F}$ and $^{*} F^{\mu\nu}$ is the dual field.

The Lagrangian density we adopt in order to describe the NED \cite{fan} is  given by
\begin{equation} \label{ned}
    \mathcal{L}(F) = \frac{4 a}{\alpha} \frac{(\alpha F)^{\frac{b+3}{4}}}{[1+(\alpha F)^{\frac{b}{4}}]^{1 + \frac{a}{b}}}
\end{equation}
where $a>0$ is a dimensionless constant which characterizes the strength of nonlinearity of the electrodynamics field, $b$ is extra a dimensionless parameter and $\alpha>0$ is constant parameter ($[\alpha]=L^2$).

In case with magnetic charges, i.e., $\phi(r)=0$, we can find different regular black hole solutions. For instance, if $a=3$ and $b=2$ in Eq.(\ref{ned}) we find the Bardeen solution \cite{bardeen}, but if $a=b$, so we have the Hayward solution [\cite{hayward}].
For a weak field strength $F$, we have that the Lagrangian density (\ref{ned}) is given by $\mathcal{L}(F) \sim \alpha^{\frac{b-1}{4}} F^{\frac{b+3}{4}}$. Therefore, only for the critical value $b=1$, it is possible to recover the Maxwellian limit. Furthermore, as noted by Ref.\cite{fan}, the regularity of black hole solutions found for $a=b$ happens when $a \geq 3$. So we choose $a=3$ in Eq. (\ref{ned} ) which leads to Hayward black hole solution [\cite{fan2}]. 
Besides, there is a class of solutions that generate a lot of interest, because in the weak field limit, the Lagrangian density (\ref{ned}) is the same as Maxwell Lagrangian, this class is reached when $a=3$ and $b=1$ \cite{pablos}, and is known as New Class. 

In this article, we study only the corrections due to the presence of the magnetic monopole $Q_m$.
%leaving the contributions due to the electric charge as a perspective. Or even the contributions of both, even knowing the already known difficulties of finding black hole solutions with dyonic charges.
The solutions of magnetically charged black holes ($\Lambda_0 =0$) in the NED for a Einstein-Hilbert gravitational action are given by \cite{maluf}
\begin{align} \label{metricned} 
   ds^2 = - &\bigg( 1 - \frac{2 M G r^{a-1}}{(q^{b}+r^{b})^{\frac{a}{b}}} \bigg)dt^2 + \bigg( 1 - \frac{2 M G r^{a-1}}{(q^{b}+r^{b})^{\frac{a}{b}}} \bigg)^{-1} dr^2 + r^2 d\theta^2 + r^2 sin^2 \theta d\phi^2,
\end{align}
where $q$ is the magnetic monopole and it relates to the total magnetic charge through $Q_m = \frac{q^2}{\sqrt{2\alpha}}$. Furthermore, $M$ is a gravitational mass (or electromagnetically induced mass), where . noted that it is necessary to make an equivalence between the gravitational mass and a kind of magnetically induced mass, i.e., $M G = q^3/ \alpha$, to obtain a correct interpretation of the parameters in the solution \cite{bobir}. Here, we are interested in study how the modified Einstein cubic gravity modifies the magnetically charged solutions.

%%%%%%%%%%%%%%%%%%%%%%%%%%%%%%%%%%%%%%%%%%%%%%%%%%%%%%%%%%%%%%%%%%%%%%%%%%%%%%%%%%%%%%%%%%%%%%%%%%%%%%%%%%%%%%%%%%%%%%%%%%%%%%%%%%%%%%%%%%%%%%%%%%%%%%55

\section{Spherically symmetric solution in (3+1) dimensions}
\label{s3}

In this section, we obtain new regular black holes solutions with spherical and static symmetry in ECG. Then, the effects of the ECG and the nonlinear electrodynamics on the spacetime geometry are investigated.

As shown in Ref.\cite{pablos}, it is possible to simplify the problem by considering the ansantz
\begin{equation} \label{metric}
    ds^2 = - N(r)^2 f(r)dt^2 + \frac{dr^2}{f(r)} + r^2 d\theta^2 + r^2 sin^2 \theta d\phi^2.
\end{equation}
In this way, the action (\ref{action}) depend on the functions $N$ and $f$, i.e., $S[N, f]$. 
Motivated by Ref.\cite{pablos}, it is possible to show that if the variation of the action (\ref{action}) in relation to the functions $N$ and $f$ vanishes,i.e.,
\begin{equation} \label{condicao}
    \frac{\delta S[N, f]}{\delta N} =   \frac{\delta S[N, f]}{\delta f}=0,
\end{equation}
then the components $\Upsilon_{tt}=\frac{1}{\sqrt{-g}}\frac{\delta S}{\delta g^{tt}}$ and $\Upsilon_{rr}=\frac{1}{\sqrt{-g}}\frac{\delta S}{\delta g^{rr}}$ of the equation of motion also vanish. Moreover, the angular equations are ensured through the Bianchi identities, $D^{\mu}\bigg[\frac{1}{\sqrt{-g}}\frac{\delta S}{\delta g^{\mu\nu}}  \bigg]=0$ when the equations $\Upsilon_{tt}=0$ and $\Upsilon_{rr}=0$ are valid.

Substituting (\ref{metric}) in (\ref{action}), we find with the aid of integration by part when necessary (surface terms are disregarded) that
\begin{align} 
\label{reducedaction} 
    & S[N, f] = \frac{1}{8\pi G} \int dr N(r) \bigg\{ -r \bigg(f(r)-1\bigg)-2 G M r^{a } \left(q^{b }+r^{b }\right)^{-\frac{a }{b }} -\frac{\Lambda_{0}  r^3}{3} \\ \nonumber
    &  -G^2 \lambda  \bigg[-\frac{24 f(r) (f(r)-1) f'(r)}{r^2}+4 f'(r)^3+\frac{12 f'(r)^2}{r}\\ \nonumber
    & -12 f(r) \bigg(f'(r)-\frac{2 (f(r)-1)}{r}\bigg) f''(r)\bigg]  \bigg \rbrace ^{'} 
\end{align}
where $' = \frac{d}{dr}$. Note that, in the reduced action in Eq.(\ref{reducedaction}), we consider that $N'$ vanishes, as suggested in [\cite{pablos}]. From that point on, we will no longer analyze the effects of the cosmological constant on our solutions, so we will assume that $\Lambda_0=0$ for the rest of the paper. Thus, by the condition (\ref{condicao}) , we reduce the problem of find the EOM to a second-order equation for a single function $f$.

In order to facilitate the numerical analysis of Eq. (\ref{reducedaction}), we introduce the dimensionless variables given by $x \equiv \frac{r}{2 MG}$, $\Tilde{\lambda}\equiv \frac{\lambda G^2}{(MG)^4}$ and $\Tilde{q} \equiv \frac{q}{2MG}$. Thus, the action in Eq.(\ref{reducedaction}) leads to the equation 
%so that the differential equation we are going to solve is given by
\begin{align} 
\label{fulleom} \nonumber
    & x \bigg(f(x)-1\bigg)+ \frac{x^{a }}{\left(\Tilde{q}^{b }+x^{b }\right)^{\frac{a }{b }}}   + \frac{\Tilde{\lambda}}{16}  \bigg[-\frac{24 f(x) (f(x)-1)}{x^2}\frac{d f}{dx} +4 \bigg(\frac{d f}{dx}\bigg)^3 +\frac{12}{x}\bigg(\frac{d f}{dx}\bigg)^2 \\ 
    &-12 f(x) \left( \frac{d f}{dx}-\frac{2 (f(x)-1)}{x}\right) \frac{d^2 f}{dx^2} \bigg] = C
\end{align}
where $C$ is an integration constant. In order to
ensure that in the limit $\lambda =0$, we obtain the solution (\ref{metricned}), we assume that $C=0$. 

The Eq.(\ref{fulleom}) is a rather complicated non-linear equation to tackle. It is worthwhile to mention that, for $q=0$ (no magnetic charge) and $\lambda=0$ (Einstein-Hilbert action), the Eq.(\ref{fulleom}) yields to $x(f-1)=-1$, whose solution is
\begin{equation}
    f(x)=1-\frac{1}{x},
\end{equation}
which is the Schwarzchild solution, as expected. Therefore, the inclusion of the ECG terms transforms the Eq.(\ref{fulleom}) from an algebraic to a differential equation.

\begin{figure}[h] % Duas figuras lado a lado
           \includegraphics[height=4.5cm]{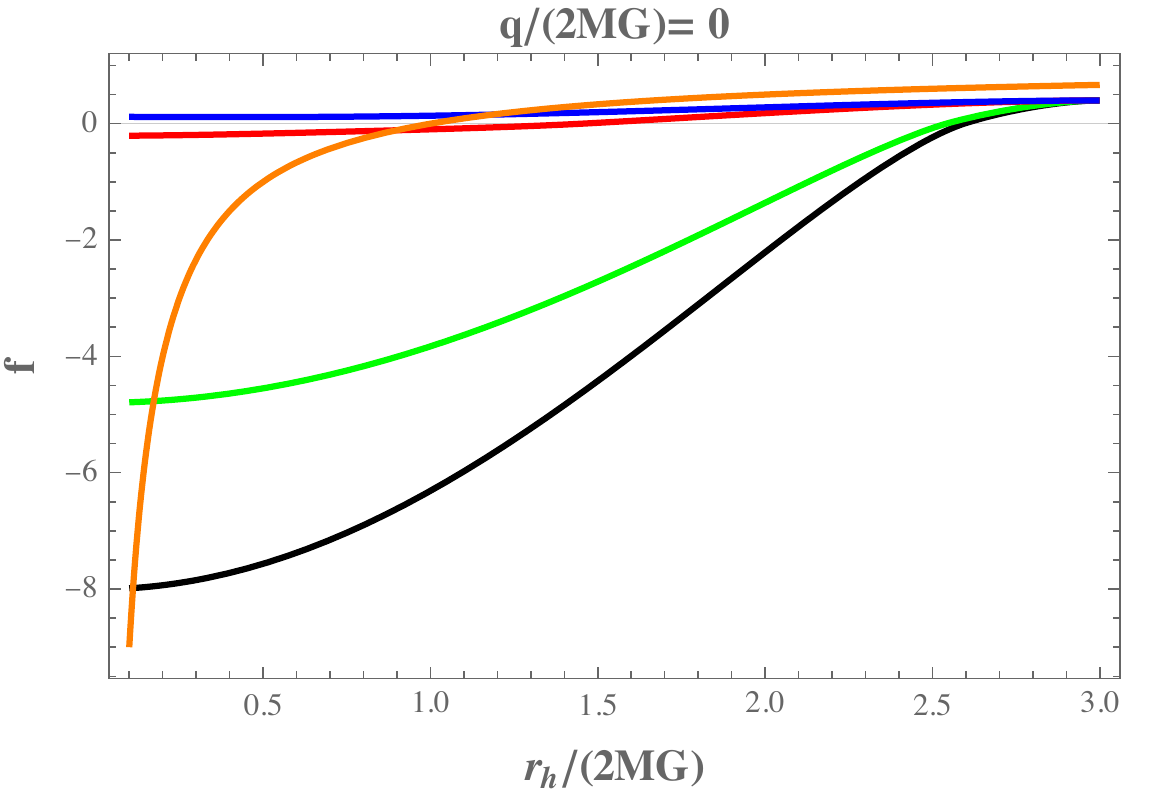}\quad
           \includegraphics[height=4.5cm]{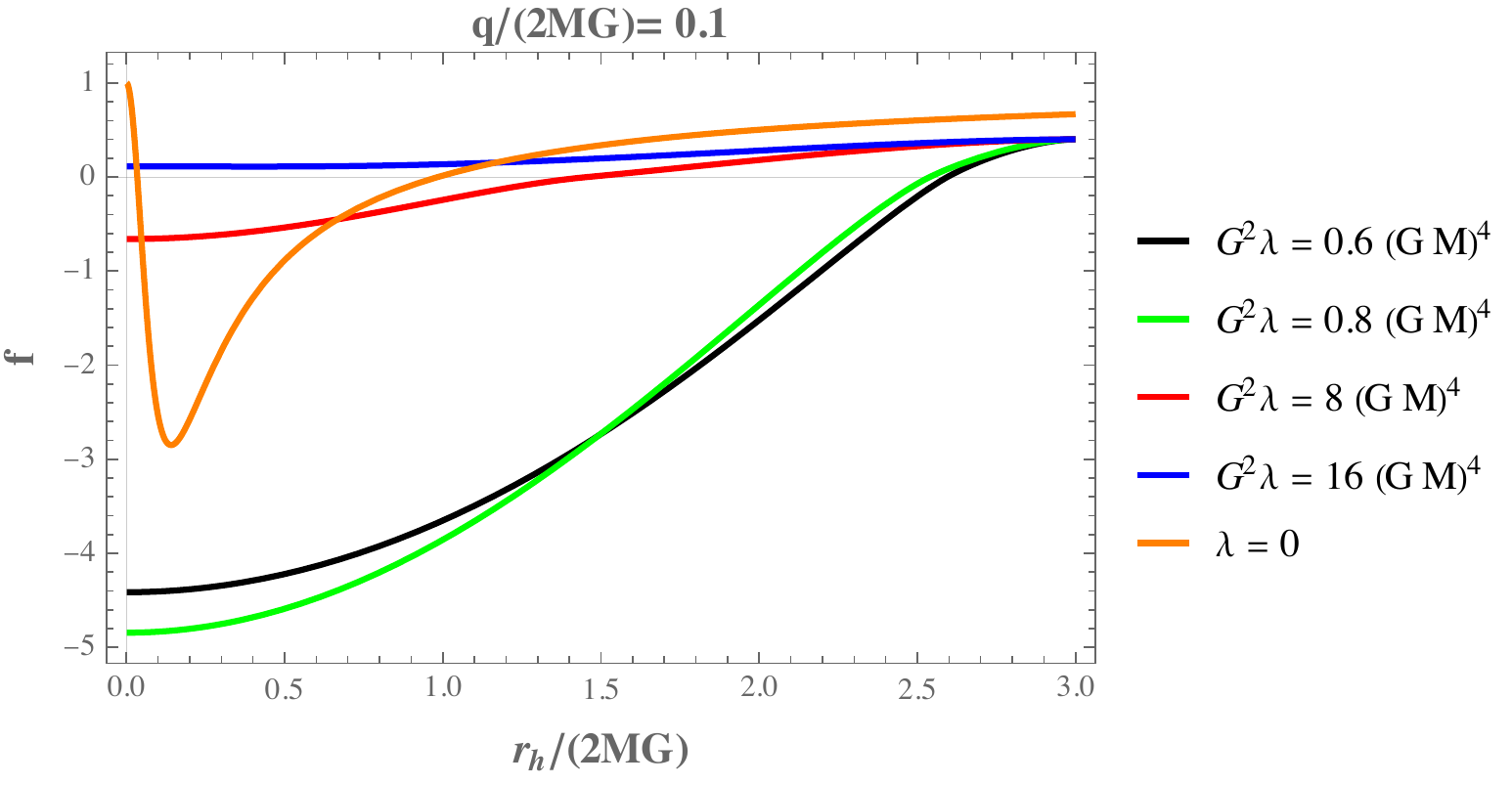}\quad
           \includegraphics[height=4.5cm]{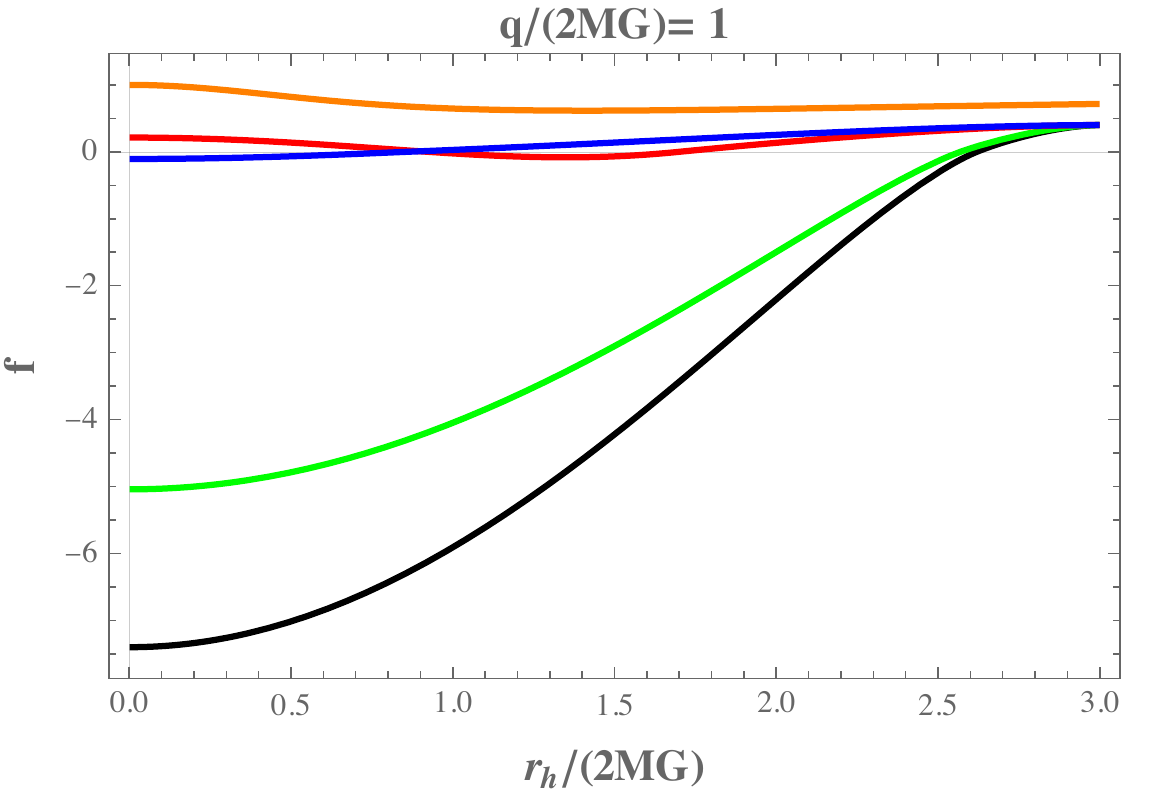}\quad
           \includegraphics[height=4.5cm]{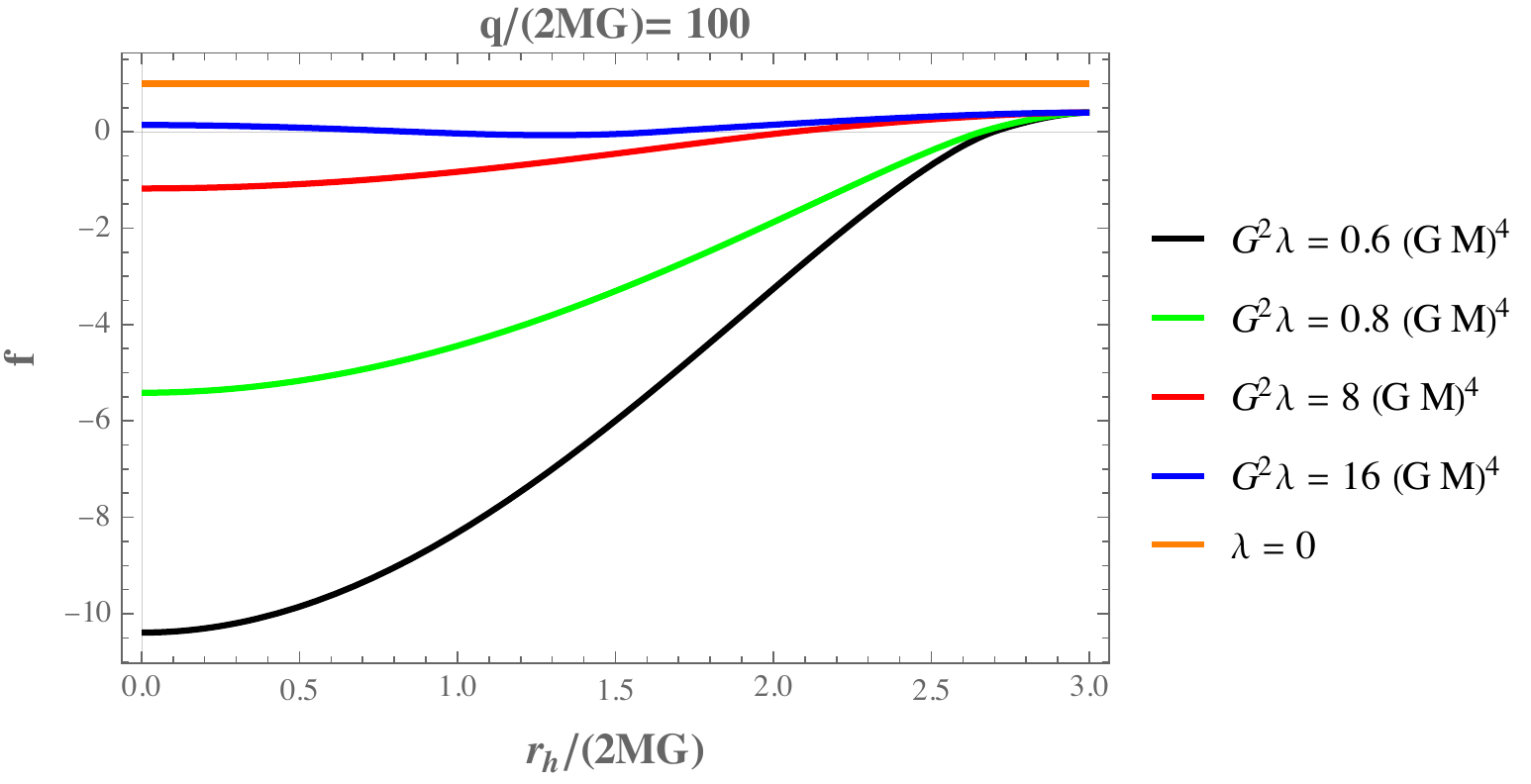}
           \caption{The metric function $f(x)$ with  null, small, unit and large $\Tilde{q}$ for various values of $\lambda$, where $x=r/2GM$.} 
          \label{fig1}
\end{figure}

We performed a numerical solution of Eq.(\ref{fulleom}) by assuming that as $x\rightarrow\infty$, then $f\rightarrow 0$ as well as $f'\rightarrow 0$. In the first Fig.(\ref{fig1}), we plotted the solutions for $\Tilde{q}=0$ and varied the values of $\lambda$.  For $G^2 \lambda=0.6(GM)^4$ (black line) and $G^2 \lambda=0.8(GM)^4$ (green line) no new event horizons appeared, we just have a shift in the size of the horizons with respect to the Schwarzschild solution (orange line), but for $G^2 \lambda=8(GM)^4$ (red line) the solution is finite at the origin and asymptotically approaches the Schwarchild solution. It is important to mention that these results agree with those found in the Ref.(\cite{pablos}). Further, the solution we found has an event horizon that is larger than Schwarzschild. Finally, we have a solution for $G^2 \lambda=16(GM)^4$ (blue line)  that has a naked singularity, although it is asymptotically flat. Thus, the ECG itself leads to a singular modified black hole solution. However, as $\lambda$ increases, the black hole  undergoes a transition into a naked singularity.

Now let us consider the influence of the magnetic charge $q$ of the Bardeen solution. Qualitatively, the numerical results for the other cases - Hayward case ($a=b=3$) and the New Class ($a=3$ and $b=1$) - are the same as those obtained for Bardeen. In Figs.(\ref{fig1}) we plotted $f(x)$ for various values of $G^2 \lambda/(2GM)^4$ with different values of $\Tilde{q}$. Note that, for $\Tilde{q}=1$ there are two event horizons in $G^2 \lambda=8(GM)^4$ (red line). On the other hand, we have only one event horizon for $G^2 \lambda=16(GM)^4$ (blue line) . For $G^2\lambda=0.8(2GM)^4$ (green line) and $G^2 \lambda=0.6(2GM)^4$ (black line) we have a singularity at the origin and one horizon.

A key feature to discuss is the presence of a singularity at the origin.
As seen from Figs. (\ref{fig1}), the Eq. (\ref{fulleom}) admits finite solutions in $r=0$. Assuming that $f(r)= \sum^{\infty}_{i=0} a_{i}r^{i} $ (returning to the original variables), where $f(0)\equiv c_0$  and $f''(0)\equiv c_2$, and substituting this power series in Eq.(\ref{fulleom}), we obtain that near the origin the solution is given by 
\begin{align} \label{origem} 
    f(r) = c_0+c_2 r^2 +\frac{ \left(48 c_2^2 G^2 \lambda -1\right)}{192 c_0 G^2 \lambda }r^4 -\frac{ \left(-16 c_2^3 G^2 \lambda  q^4+c_2 q^4+2 G M q\right)}{576 (c_0-1) c_0 G^2 \lambda  q^4}r^6
\end{align}
where we assume that $f'(0)=0$ and $c_{0,2}$ are free parameters.
The Kretschmann scalar, $K=R_{\mu\nu\rho\sigma}R^{\mu\nu\rho\sigma}$ behaves at the origin as $K\sim \frac{(c_0-1)^2}{r^4}$. Thus, the solution is singularity-free provided that $c_0 =1$. Interestingly, it turns out that the solution (\ref{origem}) is the same for the three cases of interest: Bardeen, Hayward and New class. Therefore, unlike the standard case of the GR, $\lambda=0$, the presence of the ECG might lead to black holes with singularities even with the addition of nonlinear electrodynamics field. In the next section, we employ a perturbative analysis upon a magnetic regular solution in order to probe the corrections given by the ECG.

%%%%%%%%%%%%%%%%%%%%%%%%%%%%%%%%%%%%%%%%%%%%%%%%%%%%%%%%%%%%%%%%%%%%%%%%%%%%%%%%%%%%%%%%%%%%%%%%%%%%%%%%%%%%%%%%%%%%%%%%%%55

\section{Perturbative analysis}
\label{sec4}

Even though the Eq.(\ref{fulleom}) has no analytic solution, we can understand how the corrections due to ECG can modify the NLE magnetic black hole solution (\ref{metricned}) through a perturbative analysis. Assuming a $f(r)$ function of form
\begin{equation}
\label{perturbed}
    f(r) = 1 - \frac{2 M G r^{a-1}}{(q^{b}+r^{b})^{\frac{a}{b}}} +  h(r),
\end{equation}
where $h(r)<<1$, i.e., we can linearize the field equation (\ref{fulleom}) about the magnetically charged background. 
%We linearize the differential Eq. (\ref{fulleom}) by keeping terms only to order $\epsilon$.(and setting the $\epsilon=1$ )
Considering only the linear terms in $h$, we obtain the following equation
\begin{equation} \label{h}
    h''(r)+\gamma (r) h'(r)+ \omega^2 h(r) =j(r),
\end{equation}
where the expressions for the $\gamma$, $\omega^2$ and $j$ coefficients have a rather cumbersome expression and we shown them in the Appendix A. We explore the behaviour of the solutions for the asymptotic region and at the origin.

%%%%%%%%%%%%%%%%%%%%%%%%%%%%%%%%%%%%%%%%%%%%%%%%%%%%%%%%%%%%%%%%%%%%%%%%%%%%%%%%%%%%%%%%%%%%%%%%%%%%%%%%%%%%%%%%%%%%%%%%%%%%%%%%%%%%%%%%%%%%%%%%5

\subsection{Asymptotic behavior}

Let us now examine the behaviour of the solution of Eq. (\ref{h}) for large $r$ . In this regime the leading terms of the coefficients $\gamma$ and $\omega^2$ are 
\begin{equation} \label{g1}
    \gamma(r) \approx  - \frac{2}{r}
\end{equation}
and
\begin{equation}\label{w1}
    \omega(r)^2 = - \frac{r^{3}}{72 G^{3} M \lambda}.
\end{equation}
Moreover, the NLE term can be expanded as 
\begin{equation}
 \label{nle1}
    -2 G M r^{a} \left(q^{b }+r^{b }\right)^{-\frac{a }{b }} \approx -2 G M  \bigg[1 - \frac{a}{b} (\frac{q}{r})^{b} + ... \bigg],
\end{equation}
which tends asymptotically to $-2GM$. 
Thus, the solution of the linearized Eq. (\ref{h}) for large r is given by
\begin{equation}
    f_{r \rightarrow \infty}(r) = 1 - \frac{2 M G}{r} - G^2 \lambda\bigg[ \frac{432(MG)^2 }{r^6} - \frac{736(M G )^3}{r^7} \bigg] + O\bigg(\frac{\lambda^2}{r^{11}} \bigg),
\end{equation}
%where we assume the the exponential terms decays super-exponentially and can therefore be neglected.
which has the same form as one found in Ref.\cite{pablos} for $q=0$. Therefore, the modifications given by the NLE magnetic charge are asymptotically suppressed when compared to the corrections driven by the ECG.

%%%%%%%%%%%%%%%%%%%%%%%%%%%%%%%%%%%%%%%%%%%%%%%%%%%%%%%%%%%%%%%%%%%%%%%%%%%%%%%%%%%%%%%%%%%%%%%%%%%%%%%%%%%%%%%%%%%%%%%%%%%%%%%%%%%%%%%%%%%%%%%%55

\subsection{Behavior at origin}
Now consider the Eq. (\ref{h}) for small $r$, i.e., near to the origin. 
The competition between the singular behaviour of the ECG and the regular NLE dynamics will determine the presence or the resolution of the singularity.

\subsubsection{Bardeen solution}
The Bardeen Lagrangian density ($a=3$ and $b=2$ in Eq. (\ref{ned})) has the coefficients $\gamma$ and $\omega^2$ given by
\begin{equation} \label{g2}
    \gamma(r) \approx \frac{3}{r}
\end{equation}
and 
\begin{equation} \label{w2}
    \omega(r)^2 \approx \frac{\xi_{B}^2}{
 r^2 }
\end{equation}
where $\xi_{B}^2 \equiv -8 + \frac{8 M G}{3q} -  \frac{q^{5}}{72 G^3 M \lambda}$. %Substituting the Eqs. (\ref{g2}) and (\ref{w2}) in Eq. (\ref{h}), we obtain that 
The homogeneous solution ($j(r)=0$) is given by $  h_{h}^{a=3,b=2}(r) \approx c_{1}r^{-(1+\sqrt{1-\xi_{B}^2})} + c_{2}r^{(\sqrt{1-\xi_{B}^2}-1)} $, where $c_{1}$ and $c_{2}$ are constants.For a regular at the origin, we choose $c_{1}=0$, so that
\begin{equation}
    h_{h}^{a=3,b=2}(r) \approx c_{2}r^{(\sqrt{1-\xi_{B}^2}-1)},
\end{equation}
where $\xi_{B}^2 \leq 0$.  At leading order, a particular solution for the Bardeen Lagrangian density is
\begin{equation} 
     h_{p}^{a=3,b=2}(r) \approx -\frac{ 128 G^5 M^3 \lambda }{q^9} r^2  + \frac{576 G^4  M^2 \lambda \left(3 q +G M\right)}{q^{11} } r^4  +O\left(\lambda^2,r^3\right).
\end{equation}
Thus, the general solution of Eq.(\ref{h}) for modified Bardeen solution by the ECG near the origin is given by
\begin{equation}
    h(r)_{r\rightarrow0}^{a=3,b=2}= c_{2}r^{(\sqrt{1-\xi_{B}^2}-1)} -\frac{ 128 G^5 M^3 \lambda }{q^9} r^2  + \frac{576 G^4  M^2 \lambda \left(3 q +G M\right)}{q^{11} } r^4  +O\left(\lambda^2,r^3\right). 
\end{equation}
It is worthwhile to mention the presence of a De Siter-like core near the origin, as shown by the term $\frac{ 128 G^5 M^3 \lambda }{q^9} r^2$. From Eq.(\ref{perturbed}), for $c_2 = 0$ the function $f$ satisfies $f(0)=1$, and thus, the Kretschmann scalar vanishes at the origin. Therefore, small perturbations of the Bardeen regular black hole driven by the ECG preserves the absence of the singularity at the origin. 
%%%%%%%%%%%%%%%%%%%%%%%%%%%%%%%%%%%%%%%%%%%%%%%%%%%%%%%%%%%%%%%%%%%%%%%%%%%%%%%%%%%%%%%%%%%%%%%%%%%%%%5

\subsubsection{Hayward solution}

Now let us analyze the Hayward Lagrangian density ($a=b=3$ in Eq. (\ref{ned}). For this non-Maxwellian weak field limit case, we find that the coefficients $\gamma$ and $\omega^2$ near the origin are given by
\begin{equation} \label{g3}
    \gamma(r) \approx  \frac{4}{r}
\end{equation}
and 
\begin{equation} \label{w3}
    \omega(r)^2 \approx \frac{\xi_{H}^2}{r^3}
\end{equation}
where $\xi_{H}^2 \equiv \frac{8 G M}{3}-\frac{q^6}{72 G^3 M \lambda }$. Thus, the solution for the homogeneous perturbed equation in Eq.(\ref{perturbed}) is given by 
\begin{align}
   & h_h ^{a=b=3} (r) = 2 \xi_{H}^3 r^{-3/2} \bigg[ -3 A I_{3}\bigg( 2 \xi_{H} r^{-1/2} \bigg) + B K_{3}\bigg( 2 \xi_{H} r^{-1/2}\bigg) \bigg]  ,
\end{align}
where $A,B$  are constants and $I_{\nu}(x)$ and $K_{\nu}(x)$ are the modified Bessel functions of the first and second kinds, respectively. To leading order in small r, this can be expanded as
\begin{equation}
     h_h ^{a=b=3} (r) \approx c_{2} \sqrt{\pi \xi_{H}^5} r^{-5/4} exp[{-\xi_{H} r^{-\frac{1}{2}}}],
\end{equation}
where $A=0$ to ensure that the solution is regular at the origin. Finally, we have that the particular solution linear in $\lambda$ is given by
\begin{equation}
    h_p ^{a=b=3} (r) \approx -\frac{128 G^5  M^3 \lambda  }{q^9}r^2 + \frac{384 G^5 M^3 \lambda }{q^{12}} r^5 +O\left(\lambda^2,r^6\right).
\end{equation}
Thus, the general solution of Eq.(\ref{h}) for Hayward near the origin is given by
\begin{equation}
    h(r)_{r\rightarrow0}^{a=b=3}= c_{2} \sqrt{\pi \xi_{H}^5} r^{-5/4} exp[{-\xi_{H} r^{-\frac{1}{2}}}] -\frac{128 G^5  M^3 \lambda  }{q^9}r^2 + \frac{384 G^5 M^3 \lambda }{q^{12}} r^5 +O\left(\lambda^2,r^6\right). 
\end{equation}
%Note that the exponential term decays super-exponentially and can therefore be neglected.
For a regular black hole solution, we can consider the particular choice $c_2 =0$.
%%%%%%%%%%%%%%%%%%%%%%%%%%%%%%%%%%%%%%%%%%%%%%%%%%%%%%

\subsubsection{New class}

As the last possible configuration for the Lagrangian (\ref{ned}), we adopt a new class of solutions obtained by Ref.[\cite{fan}], whose weak field limit leads to Maxwell Lagrangian. This new class is defined when $b=1$. As already mentioned, it is necessary that $a \geq 3$ to reach the regularity of the solutions at the origin. Thus again we choose the case when $a=3$. The coefficients $\gamma$ and $\omega^2$ for this choice are
\begin{equation} \label{g4}
    \gamma(r) \approx  \frac{2}{r}
\end{equation}
and 
\begin{equation} \label{w4}
    \omega(r)^2 \approx - \frac{6}{r^2},
\end{equation}
so that we have that the homogeneous solution is given by $h_h ^{b=1, a=3} (r) \approx \Tilde{c} r^2 + \frac{\Tilde{d}}{r^3}$, where $\Tilde{c}$ and $\Tilde{d}$ are constants. For $\Tilde{d}=0$ we obtain a regular homogeneous solution in the origin given by
\begin{equation}
    h_h ^{a=3,b=1} (r) \approx \Tilde{c} r^2.
\end{equation}
On the other hand, the particular solution to first order in $\lambda$ is given by 
\begin{align}
   &h_p ^{a=3,b=1}  (r) \approx \frac{16 G^4 M^2  (81q - 8 G M)\lambda}{q^9} r^2 + \frac{576 G^4 M^2(2 GM-21 q)\lambda}{q^{10}} r^3 +O\left(\lambda^2,r^4\right).
\end{align}
Thus, the general regular solution of Eq.(\ref{h}) for New Class near the origin is given by
\begin{equation}
    h(r)_{r\rightarrow0}^{a=3,b=1} =\Tilde{c} r^2 +  \frac{16 G^4 M^2  (81q - 8 G M)\lambda}{q^9} r^2 + \frac{576 G^4 M^2(2 GM-21 q)\lambda}{q^{10}} r^3 +O\left(\lambda^2,r^4\right). 
\end{equation}
Therefore, for all nonlinear electrodynamics solutions studied (Bardeen, Hayward, New class), the perturbations given by the ECG still yield to regular black holes. Moreover, around the origin, the metric exhibits a De Sitter-like behaviour whose local cosmological constant is proportional to the magnetic charge, the mass and the ECG constant $\lambda$.
%It is worth remembering that all our calculations are made assuming that $\lambda \geq 0$.

%%%%%%%%%%%%%%%%%%%%%%%%%%%%%%%%%%%%%%%%%%%%%%%%%%%%%%%%%%%%%%%%%%%%%%%%%%%%%%%%%%%%%%%%%%%%%%%%%%%%%%%%%%%%%%%%

\subsection{Horizon}

After we studied the modifications performed by the ECG upon the regular solutions at the origin and the asymptotic region, let us now explore the effects of the ECG on the black hole horizon.

We assume that the solution of the EOM (\ref{fulleom}) vanishes in $r_h$, i.e., $f(r_{h})=0$ and $f'(r_h)\geq0$. We begin by solving the field equation (\ref{fulleom}) via a series expansion near the horizon using the ansatz
\begin{equation}\label{taylor}
    f(r) = 2\kappa_g (r-r_h)+ \sum^{\infty}_{i=2} a_{i}(r-r_h)^{i} 
\end{equation}
where $\kappa_g=\frac{f'(r_h)}{2}$ is the surface gravity. 
%As seen numerically, it is possible to obtain solutions of Eq.(\ref{fulleom}) without the presence of horizons, but this only occurs for an interval of values of $\lambda$ and $q$, so that there is no problem assuming the ansatz (\ref{taylor}).That said, 
By substituting the Taylor expansion (\ref{taylor}) into Eq. (\ref{fulleom}) and solving it order by order in $(r-r_h)^i$, the two lowest-order equations are given by, respectively,
\begin{equation}\label{s1}
  r_h -\frac{2 G M r_h^{a }}{ \left(q^{b }+r_{h}^{b }\right)^{\frac{a }{b }}}-G^2 \lambda  \left(\frac{48 \kappa_{g}^2}{r_h}+32 \kappa_{g}^3\right)=0
\end{equation}
\begin{equation} \label{s2}
\frac{2 G M a q^{b } r_{h}^{a -1} }{\left(q^{b }+r_{h}^{b }\right)^{\frac{a +b }{b }}}+\frac{48 G^2 \kappa_{g}^2 \lambda }{r_{h}^2} +2 r_{h} \kappa_{g}-1=0.
\end{equation}
The higher-order coefficients $a_i$ with $i\geq3$ are determined by quite cumbersome expressions involving $\kappa_g$, $M$, $q$ and $a_2$ (free parameter).

Solving the Eq. (\ref{s2}) for $\kappa_g(r_h)$, we find that the surface gravity is given by
\begin{equation} \label{sg}
   \kappa_g = \frac{r_{h}^2  - \frac{2 M G a q^{b} r_{h}^{a+1} }{(q^{b}+ r_{h}^{b})^{\frac{a+b}{b}}} }{ r_{h}^3 \bigg(1 + \sqrt{  1 + \frac{48 G^2 \lambda}{r_{h}^4}\bigg( 1 - \frac{2 M G a q^{b} r_{h}^{a-1} }{(q^{b}+ r_{h}^{b})^{\frac{a+b}{b}}} \bigg)  }\bigg) }
\end{equation}
Note that for $\lambda=0$, we recover the standard result predicted by General Relativity. Obviously, there is another possible solution for the Eq. (\ref{s2}), however it does not reproduce the correct result in the limit case $\lambda=0$. Moreover, assuming that $\frac{q}{r_h}<<1$, the surface gravity up to first-order in $\left(\frac{q}{r_h}\right)^b$ reads 
\begin{equation}
\label{linearsurfacegravity}
\kappa_g =   \frac{1}{r_h \Big[1+\sqrt{1+\frac{48\lambda G^2}{r_{h}^{4}}}\Big]} - \frac{a M G}{r_h^2\sqrt{1+\frac{48\lambda G^2}{r_{h}^{4}}}}\bigg(\frac{q}{r_h}\bigg)^b
\end{equation}
For $q=0$, the surface gravity in Eq.(\ref{linearsurfacegravity}) leads to neutral ECG expression found in Ref.(\cite{pablos}).

By substituting the Eq.(\ref{linearsurfacegravity}) into the Eq.(\ref{s2}) and retaining the terms linear in $\left(\frac{q}{r_h}\right)^b$ leads to
%assuming that $\frac{q}{r_h}<<1$, i.e., $( q^b+r_h^b)^n \equiv r_h^{n b}\bigg[1 + n \bigg(\frac{q}{r_h}\bigg)^b + ... \bigg]$, it is possible to obtain the even the b-order in $q/r_b$ that 
\begin{equation}
\label{linearmass}
    \frac{2GM}{r_h} \approx \mu_0 \Bigg[1+ \Big[\frac{a}{b}-\frac{48 a \lambda G^2}{r_{h}^{4}\sqrt{1+\frac{48\lambda G^2}{r_h^4}}}\frac{\bigg(2 + \sqrt{1+\frac{48\lambda G^2}{r_h^4}}\bigg)}{\bigg(1 + \sqrt{1+\frac{48\lambda G^2}{r_h^4}}\bigg)^2}\Big]\left(\frac{q}{r_h}\right)^b\Bigg],
\end{equation}
where $\mu_0=\mu_0 (r_h)$ is the ECG correction given by \cite{pablos}
\begin{equation}
 \mu_0 = 1-\frac{16\lambda G^2}{r_{h}^{4}}\frac{\left(5+3\sqrt{1+\frac{48\lambda G^2}{r_{h}^{4}}}\right)}{\left(1+\sqrt{1+\frac{48\lambda G^2}{r_{h}^{4}}}\right)^3}.   
\end{equation}
The Eq.(\ref{linearmass}) allow us to relate the mass $M$ to the radius of the horizon $r_h$. For $\lambda=q=0$, we obtain the usual Schwarzchild radius, i.e., $r_h =r_s =2GM$. For $\lambda\neq 0$ and $q=0$ (neutral ECG), the $\mu_0$ factor guarantees that $r_h \geq r_s$ \cite{pablos}. On the other hand, for $\lambda\neq 0$ and $q\neq 0$, the Eq.(\ref{linearmass}) shows that the displacement of the horizon radius depends on the magnetic charge $q$ and the specific model given by $a$ and $b$.

\begin{figure}[h] 
       \includegraphics[height=4.1cm]{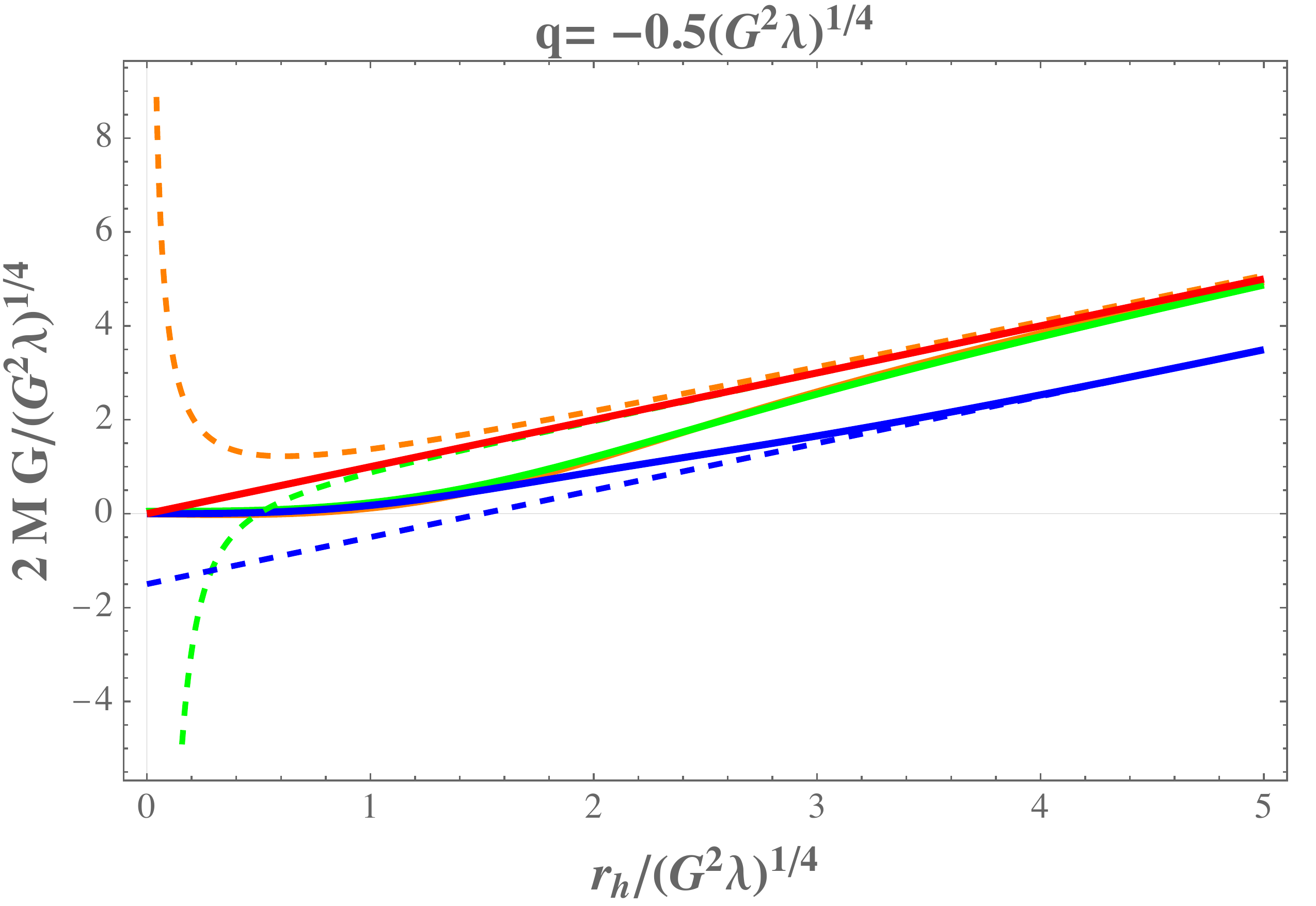}\quad
        \includegraphics[height=4.1cm]{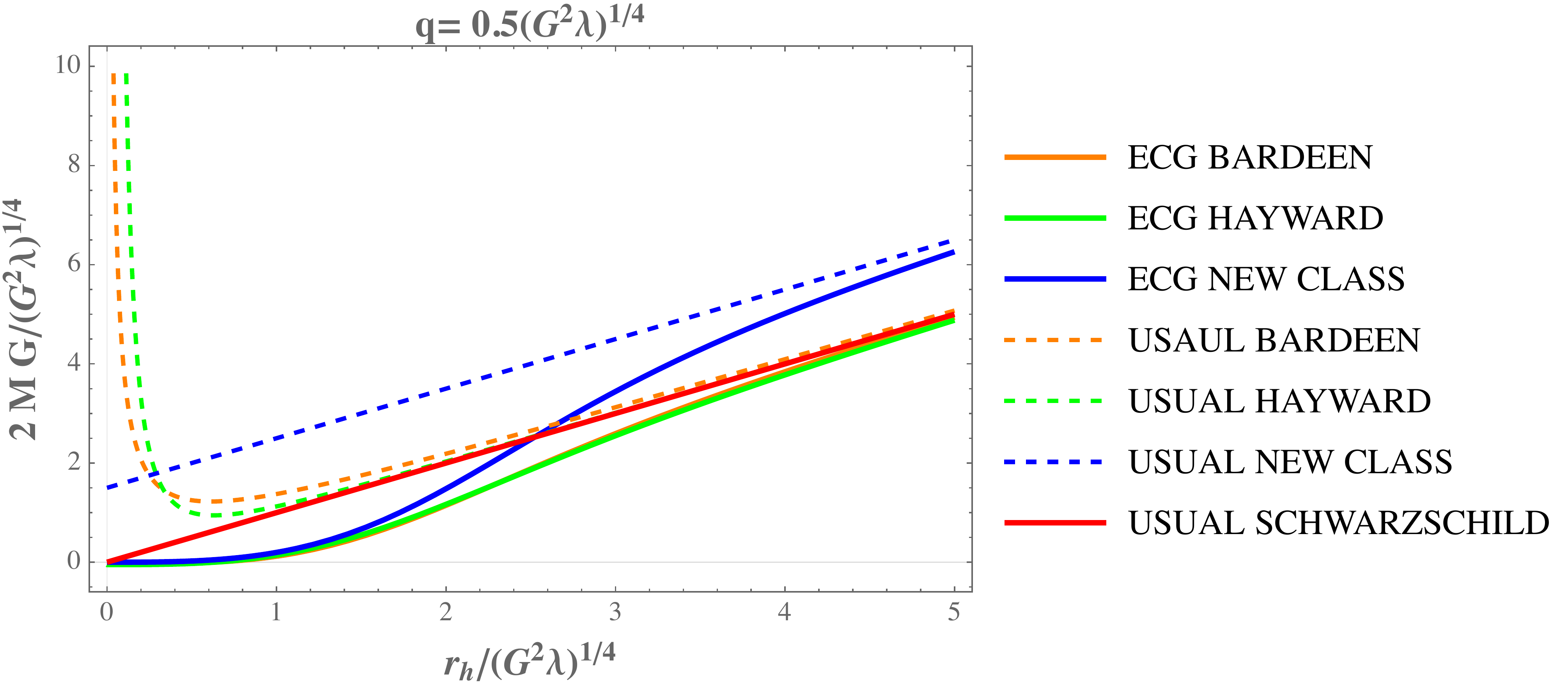}\quad
           \caption{ For $\lambda>0$, we plot $2 G \Tilde{M}$ as a function of $\Tilde{r_h}$ for different classes of Nonlinear Electrodynamics for small magnetic charge $\pm q$ compared to the horizon, where $\Tilde{M}=\frac{M}{(G^2 \lambda)^{\frac{1}{4}}}$ and  $\Tilde{r}=\frac{r}{(G^2 \lambda)^{\frac{1}{4}}}$ .} 
          \label{fig8}
\end{figure}

Indeed, as shown in the fig.(\ref{fig8}), for a small $q$ the ECG modified regular black holes have larger horizon radius for the same $M$. Nevertheless, for a bigger $q$ the ECG Hayward black hole has a smaller horizon radius. Note that the ECG does not produce new horizons.

Similarly, it is possible to obtain an expression for $q(r_h)$ for a given $M$ from Eq. (\ref{s1}) assuming the $q<<2MG$. It is worth noting that to obtain Eq.(\ref{linearmass}) we compare the magnetic charge to the event horizon, on the other hand to get a $q(r_h)$ expression, we assume that the charge is much smaller than the Scharszchild radius rather than $r_h$.

\begin{figure}[!ht] 
      \includegraphics[height=4.1cm]{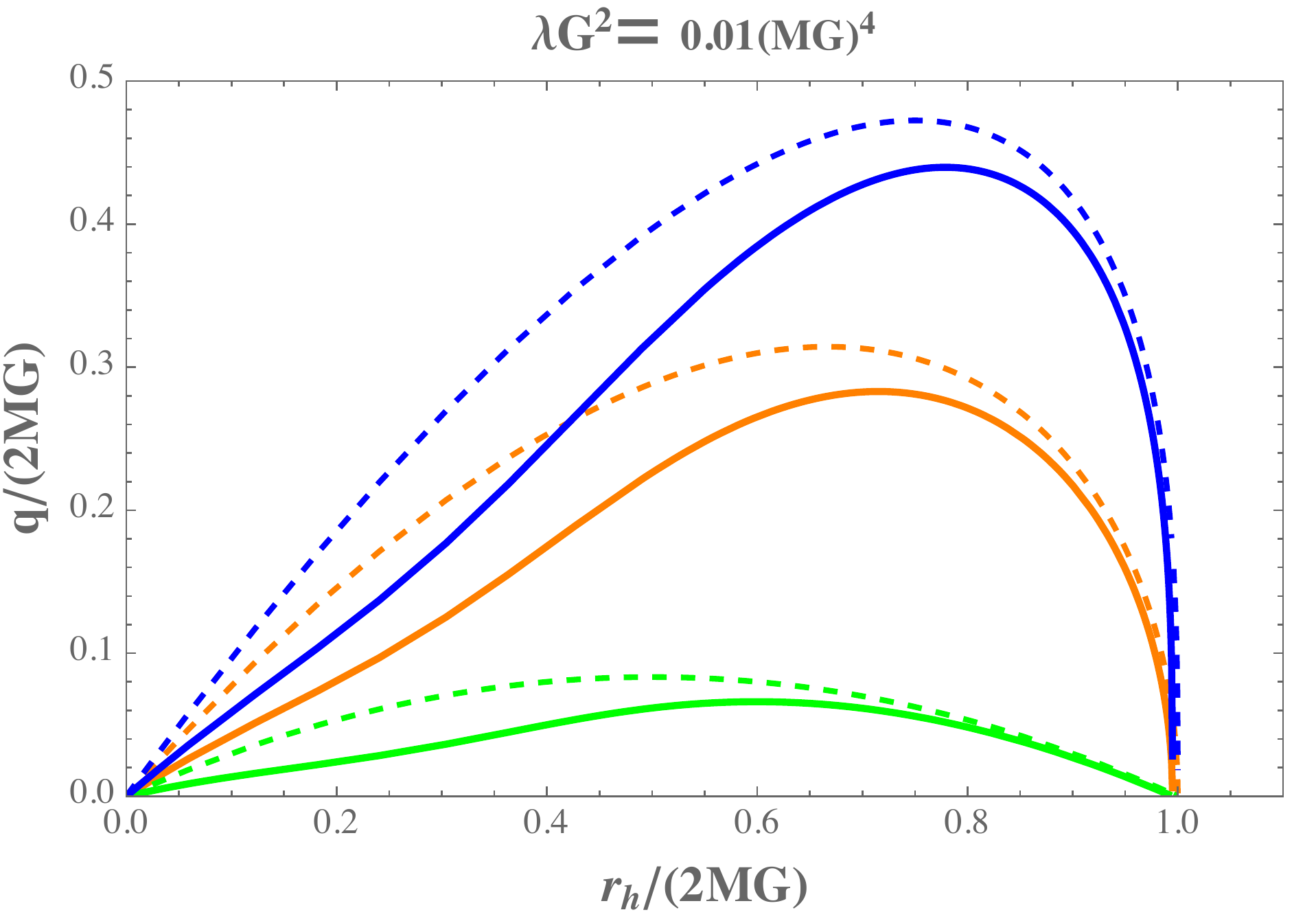}\quad
      \includegraphics[height=4.1cm]{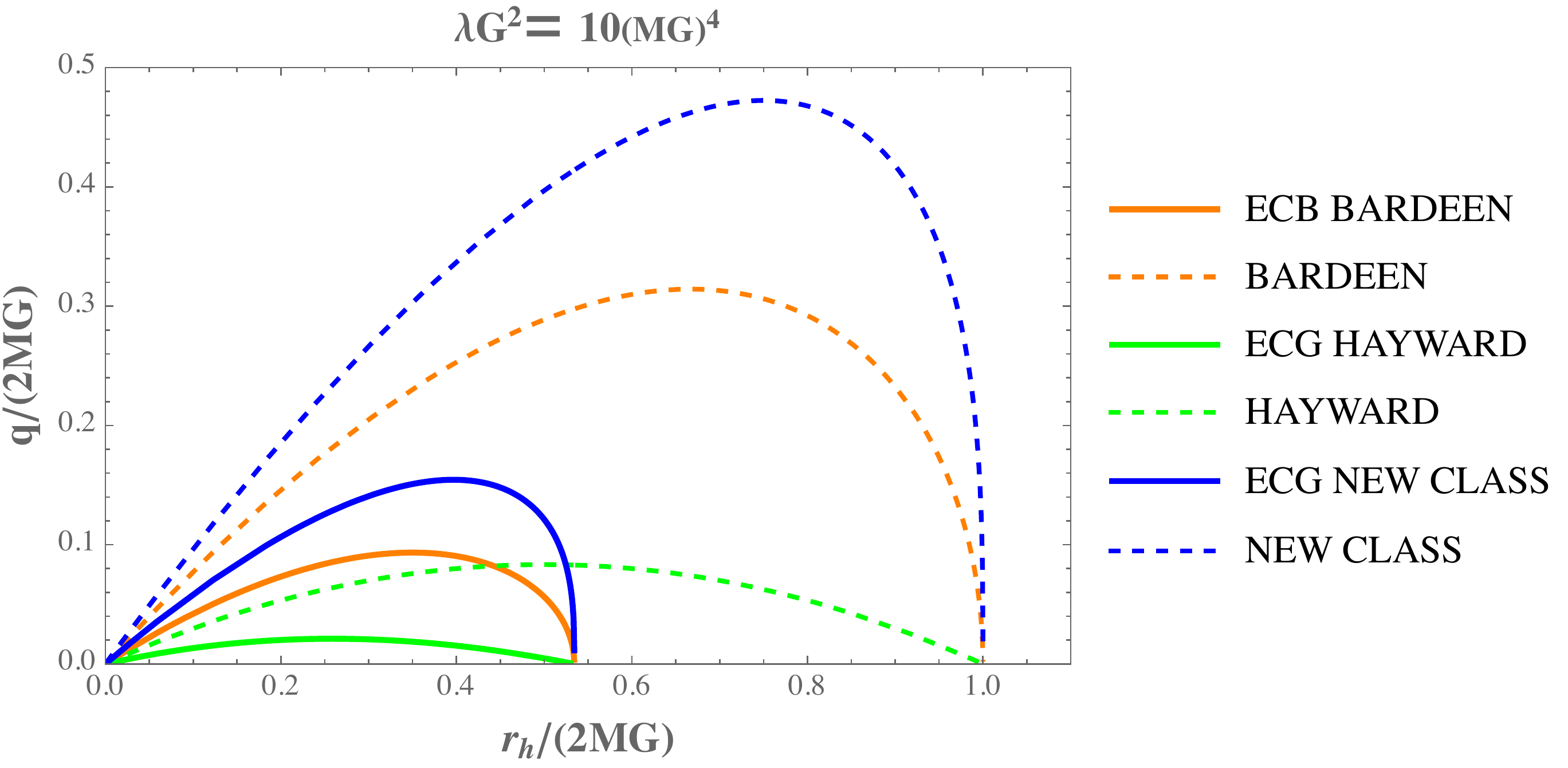}\quad
    \caption{We plot the normalized magnetic charge as a function of the horizon for small and large $\lambda$, assuming that $q<<2MG$.}  \label{carga}
\end{figure}

We plot the normalized magnetic charge $\Tilde{q}$ as a function of the normalized horizon $x$ for large and small values of $\lambda$ in Fig. \ref{carga}. 

%%%%%%%%%%%%%%%%%%%%%%%%%%%%%%%%%%%%%%%%%%%%%%%%%%%%%%%%%%%%%%%%%%%%%%%%%%%%%%%%%%%%%%%%%%%%%%%%%%%%%%%%%%%%%%%%%%%%%%%%%%%%%%%%%%%%%%%%%%%%%%%%%%%%%%%%%%%%%%%55

\section{Thermodynamics}
\label{sec5}

As discussed in the previous section, the ECG 
provides interesting new modifications for the regular black hole event horizon. Since the event horizon features are related to the black hole thermodynamics, in this section we explore the effects of the ECG on the black hole thermodynamic functions.

Let us start with the black hole entropy. It is possible to evaluate the entropy of these black hole solutions by Wald’s formula \cite{w1,w2}. Using the Eq. (\ref{sg}) this formula and considering only the contributions due to cubic gravity for $q<<r_h$, the Wald entropy up at leading order is given by
% \begin{align} \label{entr} 
%    \textbf{S} = \frac{\pi  r_h^2}{G} -\frac{48 \pi  G \lambda  %\left(r_h- \frac{2 a G M r_h^a q^b}{\left(q^b+r_h^b\right)^{\frac{a+b}{b}}} \right)^2}{r_h ^4 \left(\sqrt{\frac{48 G^2 \lambda  \left(1- \frac{2 a G M r_h^{a-1} q^b}{\left(q^b+r_h^b\right)^{\frac{a+b}{b}}} \right)}{r_h^4}+1}+1\right)^2}  \left(\frac{2 r_h \left(\sqrt{\frac{48 G^2 \lambda  \left(1-\frac{2 a G M r_h^{a-1} q^b}{\left(q^b+r_h^b\right)^{\frac{a+b}{b}}} \right)}{r_h^4}+1}+1\right)}{r_h-\frac{2 a G M r_h^a q^b}{\left(q^b+r_h^b\right)^{\frac{a+b}{b}}} }+1\right)
 %\end{align}
 \begin{align}
   &  \textbf{S} \approx \frac{\pi  r_h^2}{G}\bigg[ 1 -\frac{48   G^2 \lambda \bigg( 3+2\sqrt{1+\frac{48 G^2 \lambda }{r_h^4}} \bigg) }{ r_h^4\left(1+\sqrt{1+\frac{48 G^2 \lambda }{r_h^4}}\right)^2} \\ \nonumber
   &- \frac{2 a M G \bigg(r_h^4\bigg(-1 + \sqrt{1+\frac{48 G^2 \lambda }{r_h^4}}\bigg) - 48 G^2 \lambda\bigg(1+\sqrt{1+\frac{48 G^2 \lambda }{r_h^4}}\bigg) \bigg)}{r_h^5\bigg(1 + \frac{48 G^2\lambda}{r_{h}^4}\bigg)}\bigg(\frac{q}{r_h}\bigg)^b\bigg]
 \end{align}
The exact expression for entropy is shown in the appendix B \ref{entr}. For $q=0$ we recover the modified ECG entropy obtained in Ref[\cite{pablos}]. Note that both the ECG and the NLE modify the GR BH entropy formula $\textbf{S} = \frac{A}{4G}$, where $A$ is the area of the event horizon. as expected for modified theories of gravity \cite{my}.  

Moreover, we can determine the Hawking temperature associated with the ECG magnetic black hole also from Eq.(\ref{sg}) through the formula $T = \frac{\kappa_g}{2\pi}$. Thus, the Hawking temperature is given by
\begin{equation} \label{tem}
    T = \frac{r_{h}^2  - \frac{2 M G a q^{b} r_{h}^{a+1} }{(q^{b}+ r_{h}^{b})^{\frac{a+b}{b}}} }{ 2\pi r_{h}^3 \bigg(1 + \sqrt{  1 + \frac{48 G^2 \lambda}{r_{h}^4}\bigg( 1 - \frac{2 M G a q^{b} r_{h}^{a-1} }{(q^{b}+ r_{h}^{b})^{\frac{a+b}{b}}} \bigg)  }\bigg) }
\end{equation}
\begin{figure}[!ht] 
      \includegraphics[height=4.1cm]{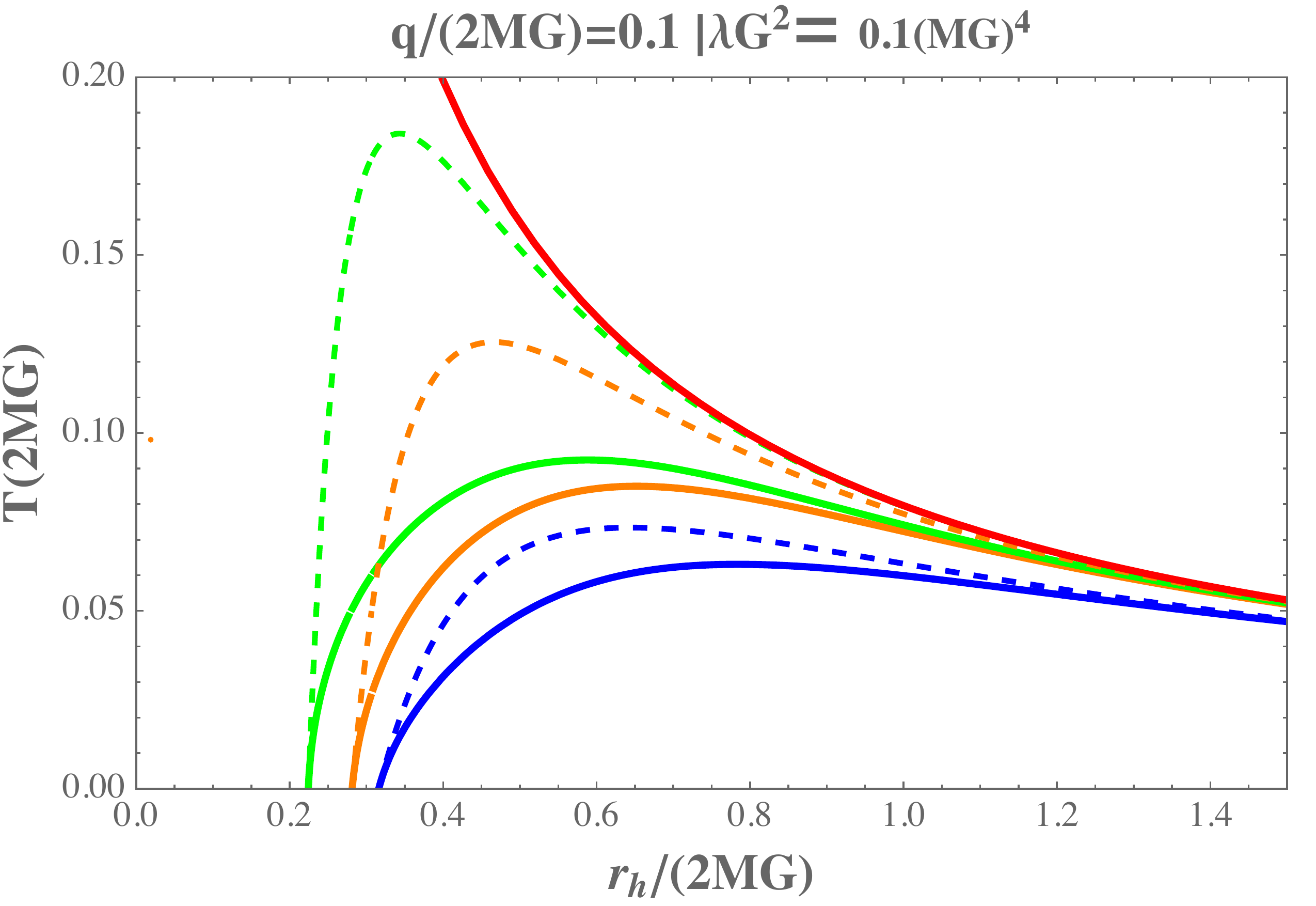}\quad
      \includegraphics[height=4.1cm]{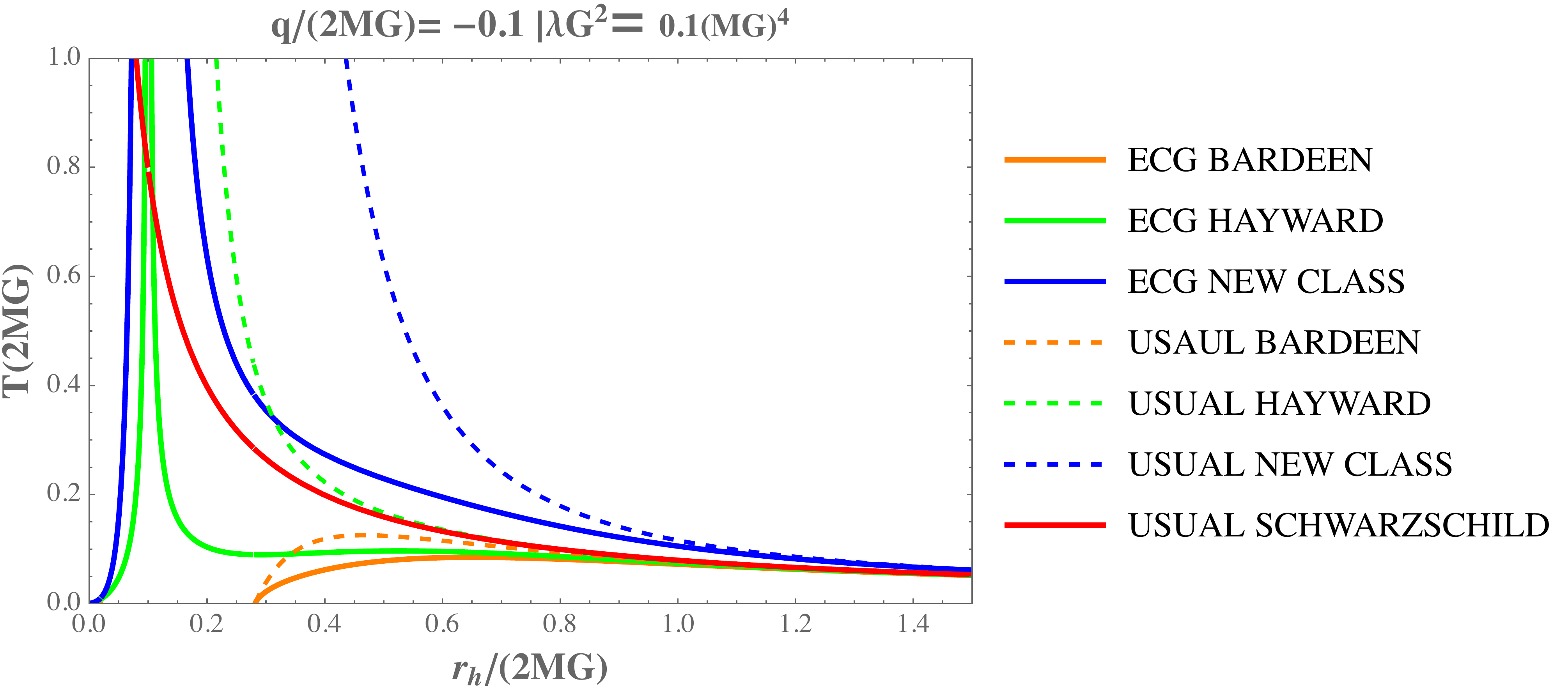}\quad
      \includegraphics[height=4.1cm]{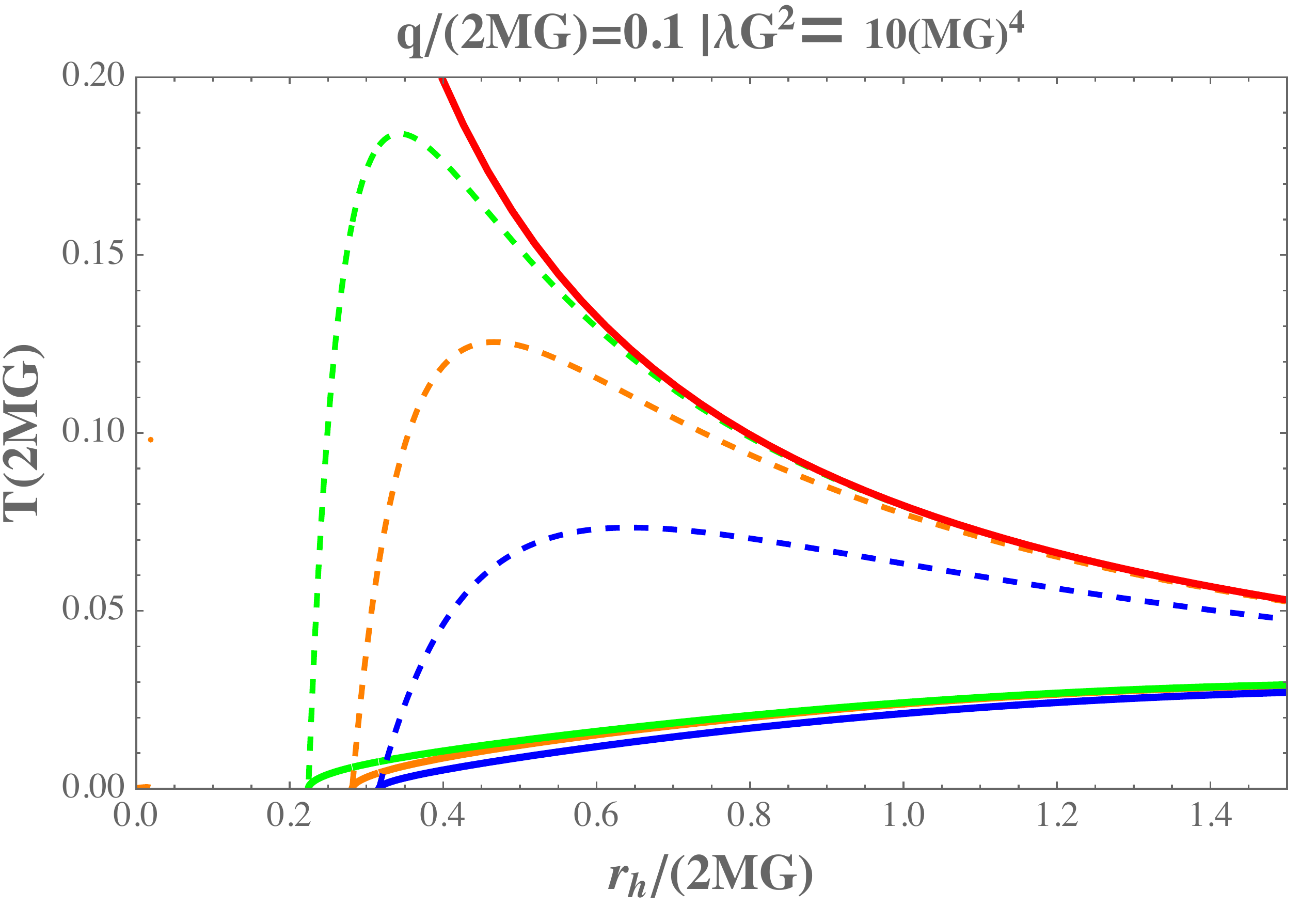}\quad
      \includegraphics[height=4.1cm]{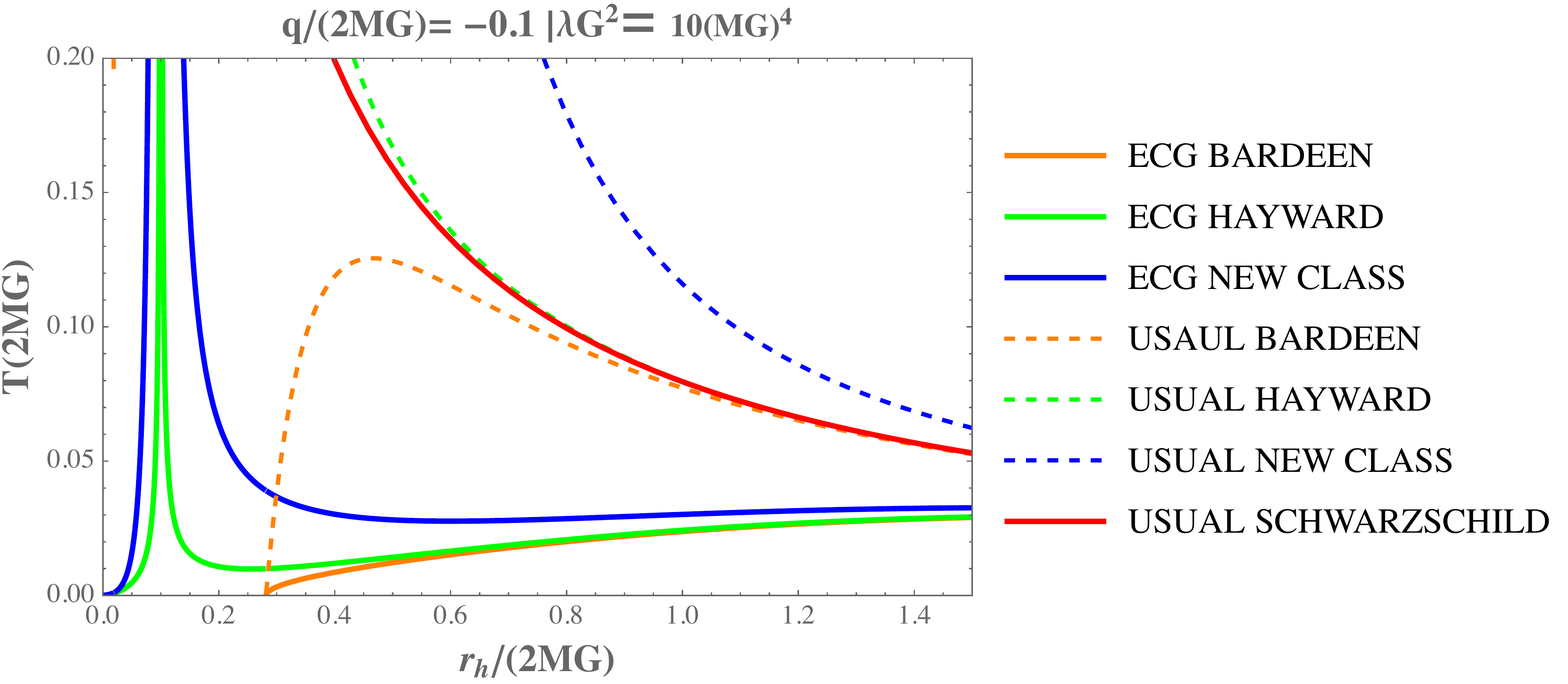}\quad
      \includegraphics[height=4.1cm]{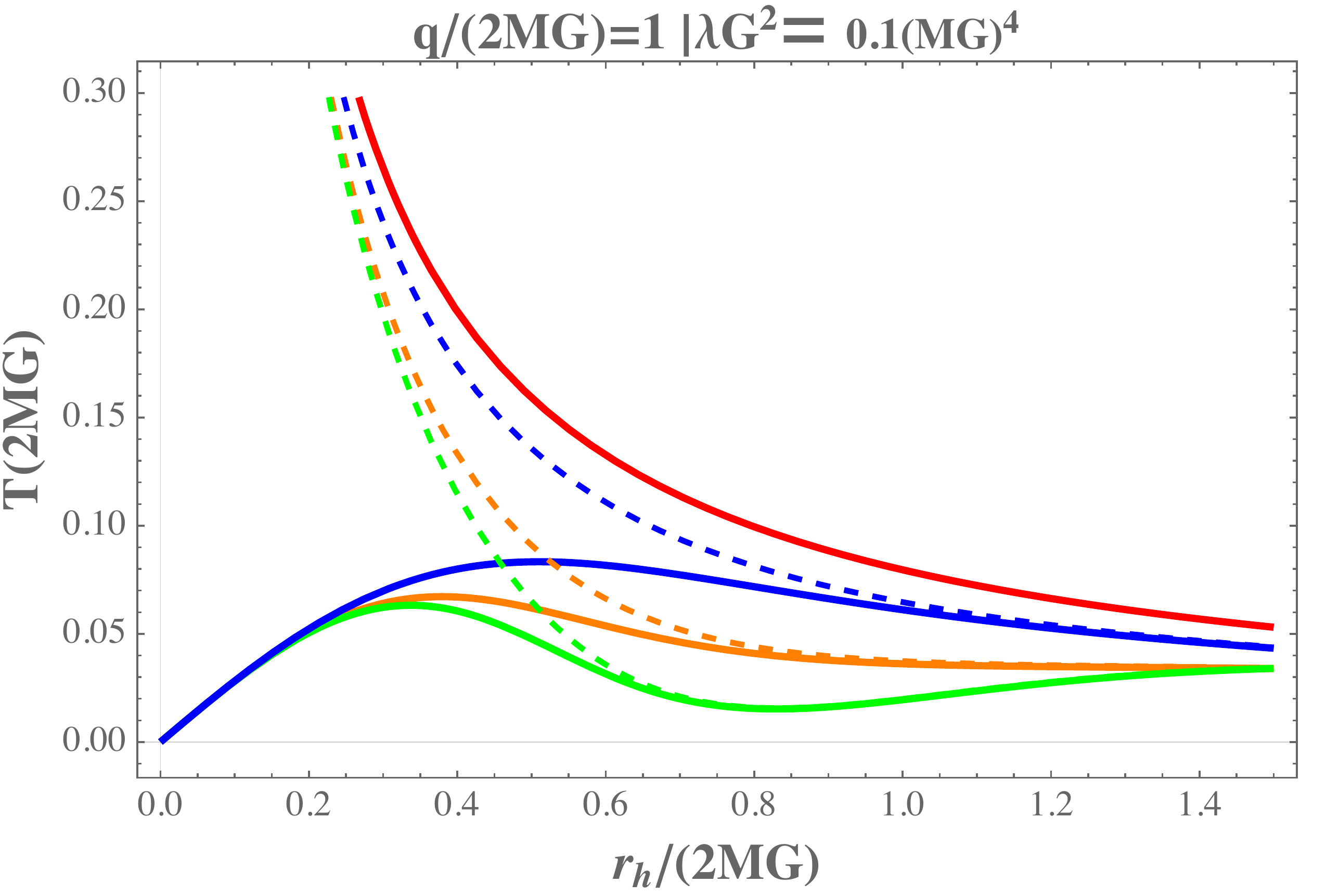}\quad
      \includegraphics[height=4.1cm]{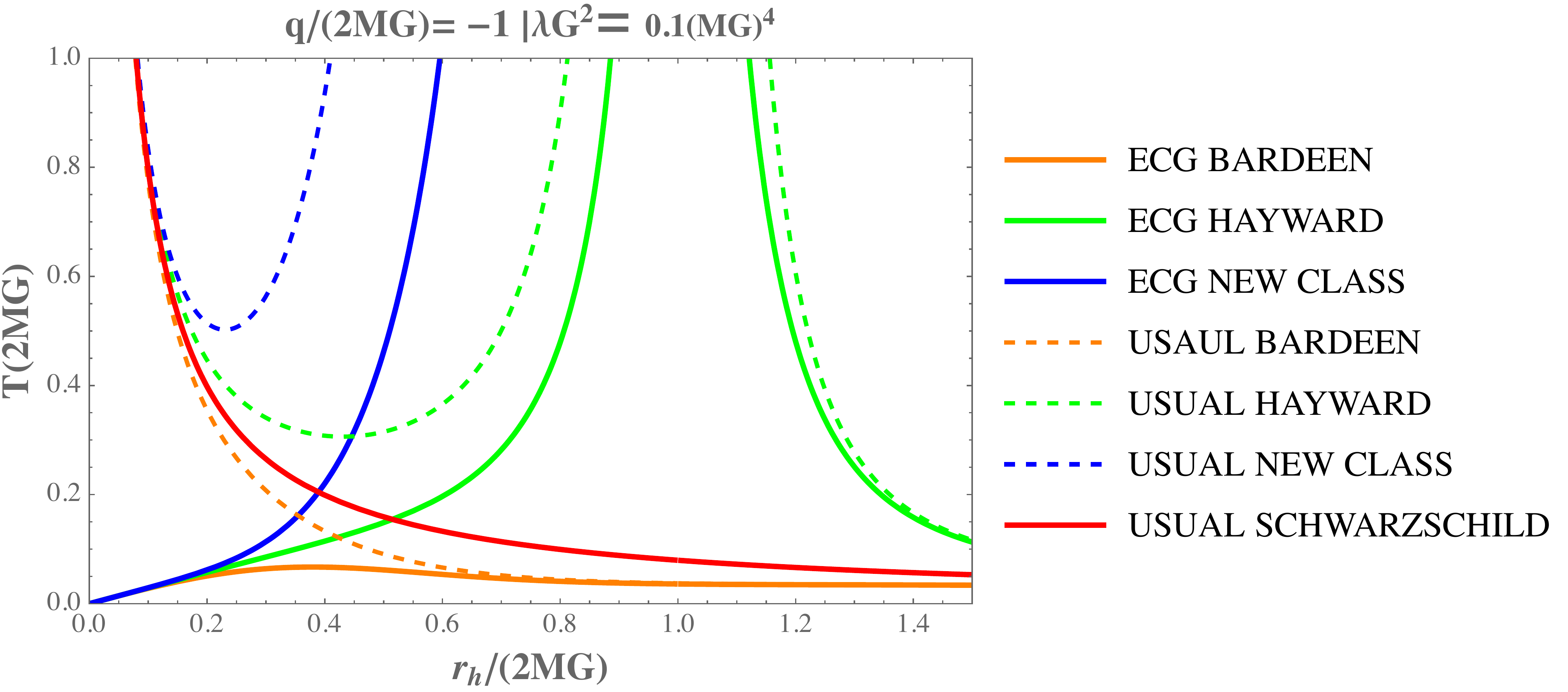}\quad
       \includegraphics[height=4.1cm]{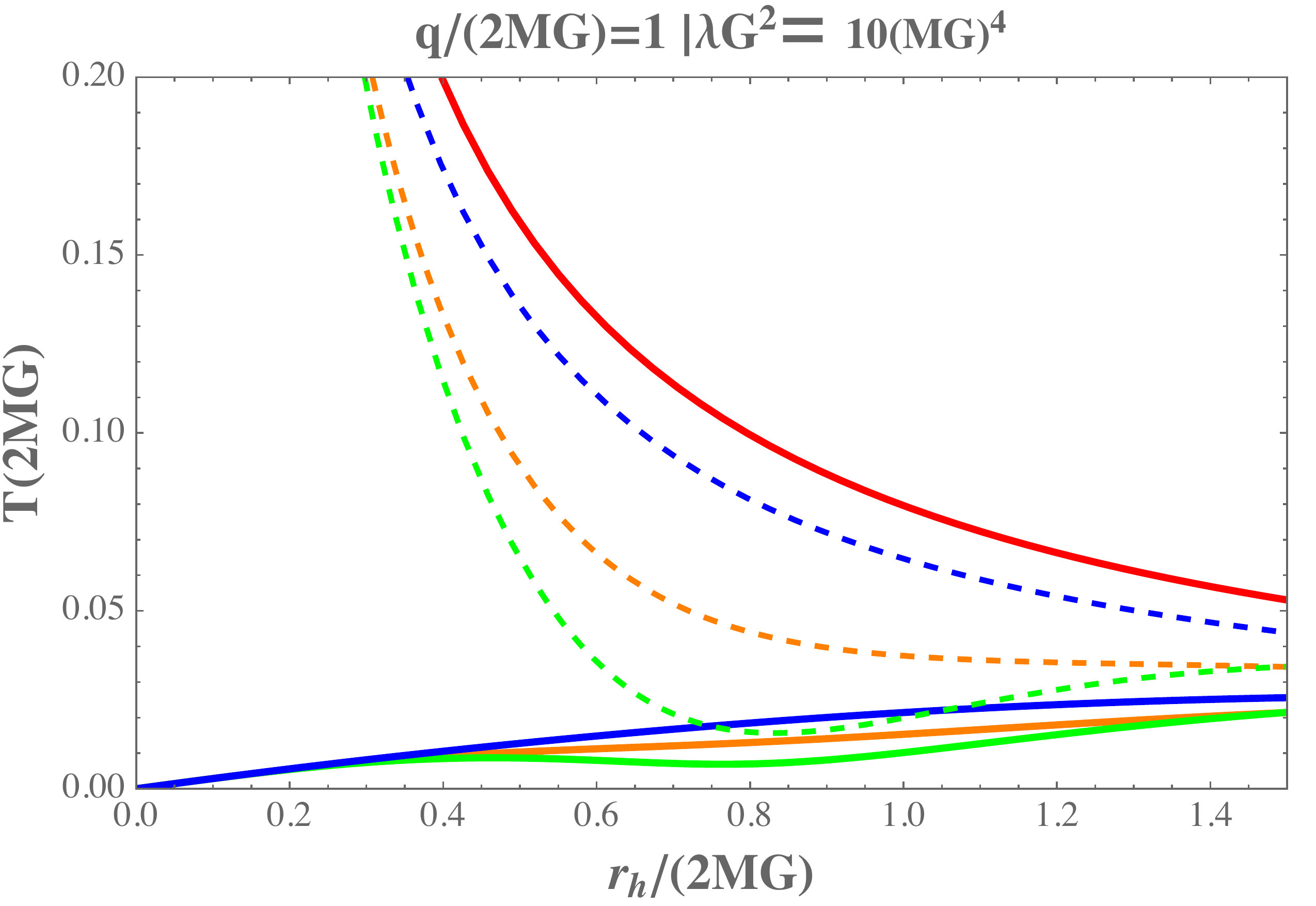}\quad
        \includegraphics[height=4.1cm]{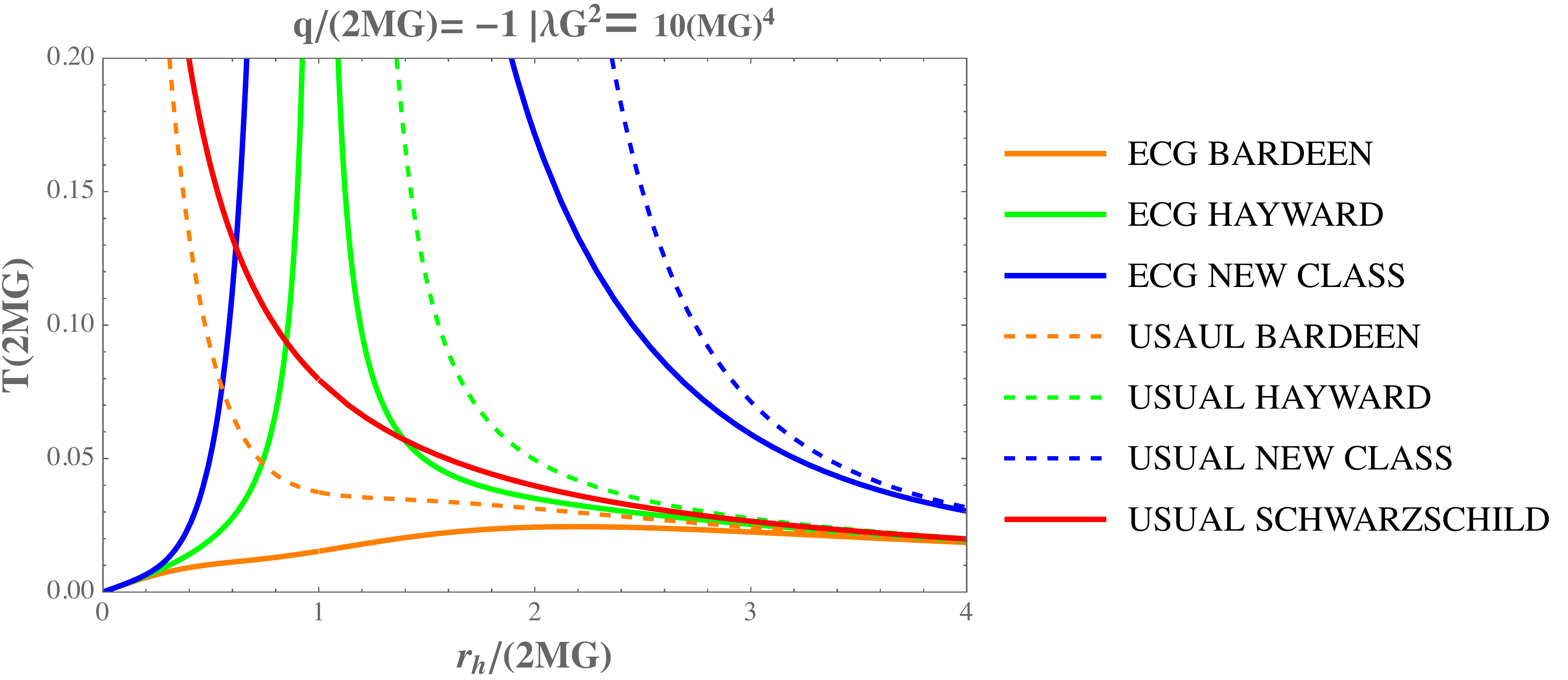}\quad
    \caption{The normalized Tempeture $\Tilde{T}= T(2 M G)$  as a function of normalized horizon radius $\Tilde{r}_h$ with small and large $\Tilde{q}$ for various $\lambda$ assuming the Bardeen ($a=3$ and $b=2$), Hayward ($a=b=3$) and New Class ($a=3$ and $b=1$) cases.}  \label{temp}
\end{figure}
For $q<<r_h$, the temperature has the form
\begin{equation} \label{teml}
    T \approx \frac{1}{2\pi r_h}\Bigg\{ \left(1+\sqrt{1+\frac{48 G^2 \lambda }{r_{h}^4}}\right)^{-1}- \frac{a GM}{ r_h\sqrt{1+\frac{48\lambda G^2}{r_{h}^{4}}}}\bigg(\frac{q}{r_h}\bigg)^b\Bigg\}
\end{equation}
Note that for $q=0$ we obtain the ECG modified Hawking temperature found in Ref.\cite{pablos}.

A noteworthy feature we found is that the Hawking temperature $T_h$ vanishes for some $r_h$ near the origin. 
The Figs.[\ref{temp}] show the temperature behavior for the Bardeen, Hayward and New Class cases as a function of the event horizon for large and small  $q$ compared to the Schawzschild radius. Unlike the Schwarzchild temperature, which diverges as $r_h \rightarrow 0$, for $q>0$ (right panel) the temperature vanishes for some $r_h \neq 0$. Since $M\neq 0$ in these configurations, the
black hole reaches a stable thermodynamic state which no longer emits, i.e., a remnant.
Moreover, as we increase the magnetic charge, the temperature for the usual regular magnetic black holes tends to diverges at the origin, whereas the modified ECG solutions lead to a vanishing Hawking temperature for $r_h =0$.

The behaviour of the Hawking temperature changes dramatically for $q<0$ (left panel). For small $q$ and without the ECG corrections, only the Bardeen black hole temperature vanishes for $r_h \neq 0$. By adding the ECG terms, the Bardeen temperature still vanishes for $r_h \neq 0$, whereas the Hayward and new class solutions undergo a phase transition. Indeed, the temperatures exhibit a Schwarzchild-like behaviour for $r_h \geq r_c$ and a decreasing ECG behaviour for $r_h \approx 0$ \cite{pablos}. As the charge increases, the Bardeen solution resembles the ECG temperature, the Hayward solution undergoes the phase transition described and the new class diverges as $r_h \rightarrow \infty$. By increasing the ECG coupling, the three regular solutions show a ECG behaviour near the origin and vanishes asymptotically.

%Notably, the temperature increases until reach its maximum in   $\frac{\partial T}{\partial r_h }=0$ for all solutions with ECG (the solid lines, except the red line, the Schwarzschild solution). %The maximum temperature, which is reached for a mass $M^{max}$, 
%has a confusing expression.  However it is possible to obtain the maximum mass if we assume that $q<<2MG$. For this approximation, considering only linear terms in $\frac{q}{2MG}$, we have that
%\begin{equation}
 %  M^{max}\approx \frac{r_h \left(\frac{q}{r_h}\right)^{-b} \left(1+\frac{48 G^2 \lambda }{r_h^4}\right) \left( \left(\sqrt{\frac{48 G^2 \lambda }{r_h^4}+1}-1\right)-\frac{16 G^2 \lambda}{r_h^4} \right)}{16 a G^3 \lambda  \left((b+2) + \frac{48 b G^2 \lambda}{r_h^4}\right)}
%\end{equation}

\begin{figure}[!ht] 
      \includegraphics[height=4.1cm]{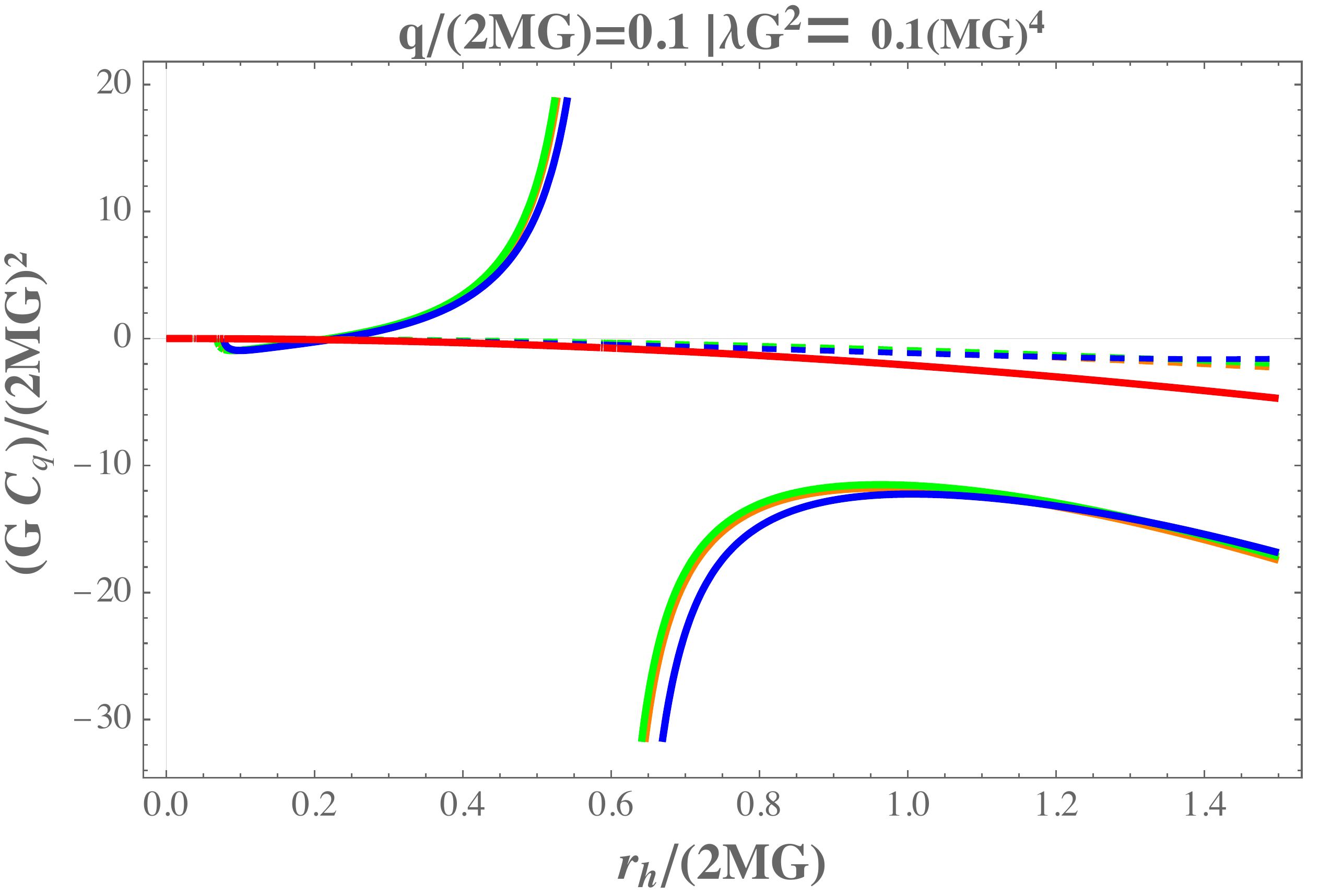}\quad
      \includegraphics[height=4.1cm]{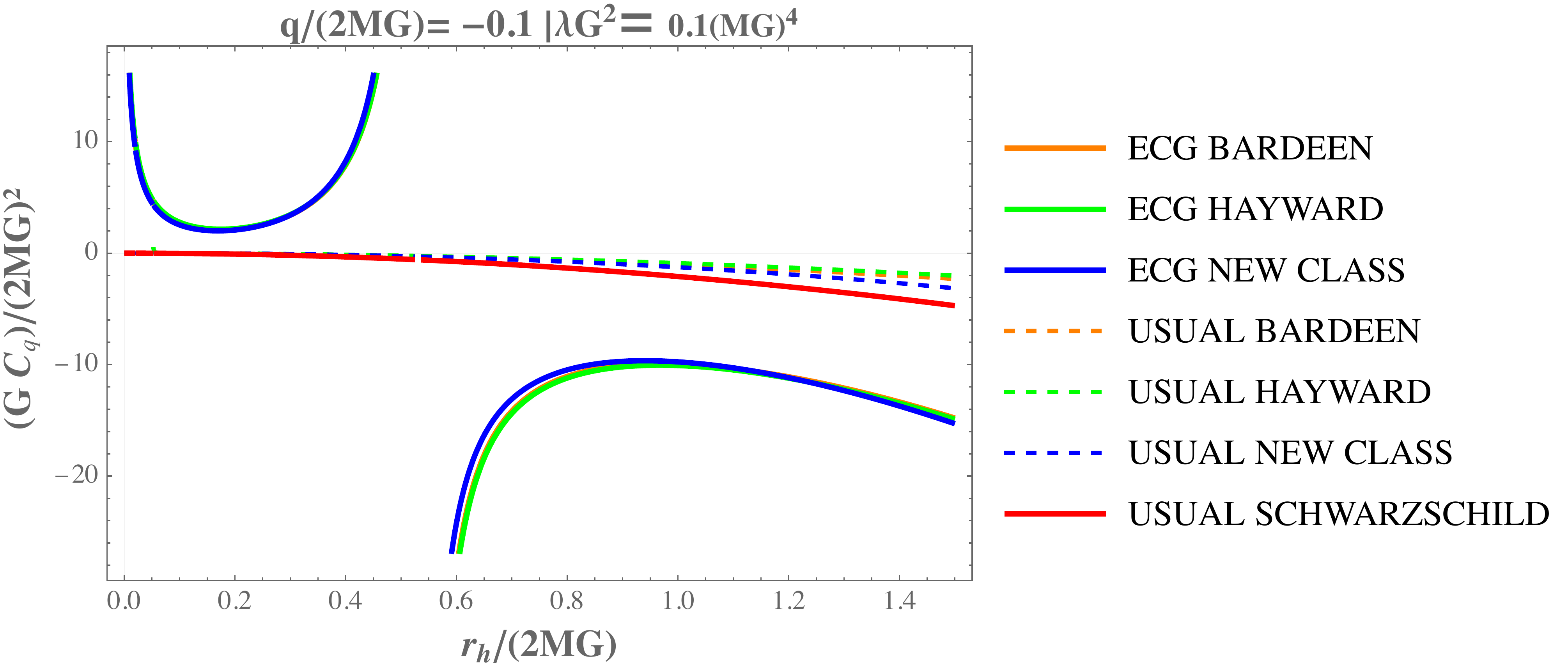}\quad
      \includegraphics[height=4.1cm]{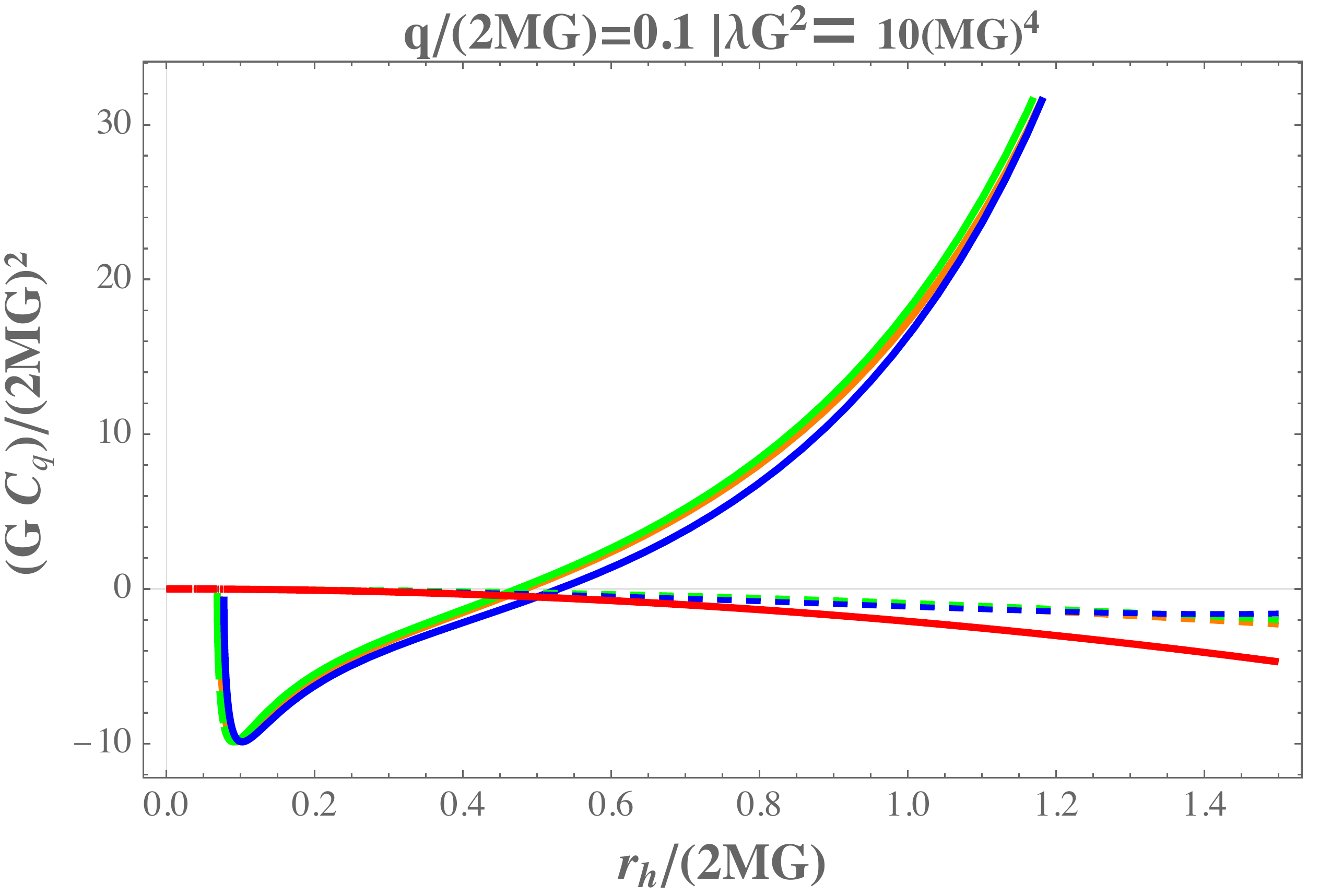}\quad
      \includegraphics[height=4.1cm]{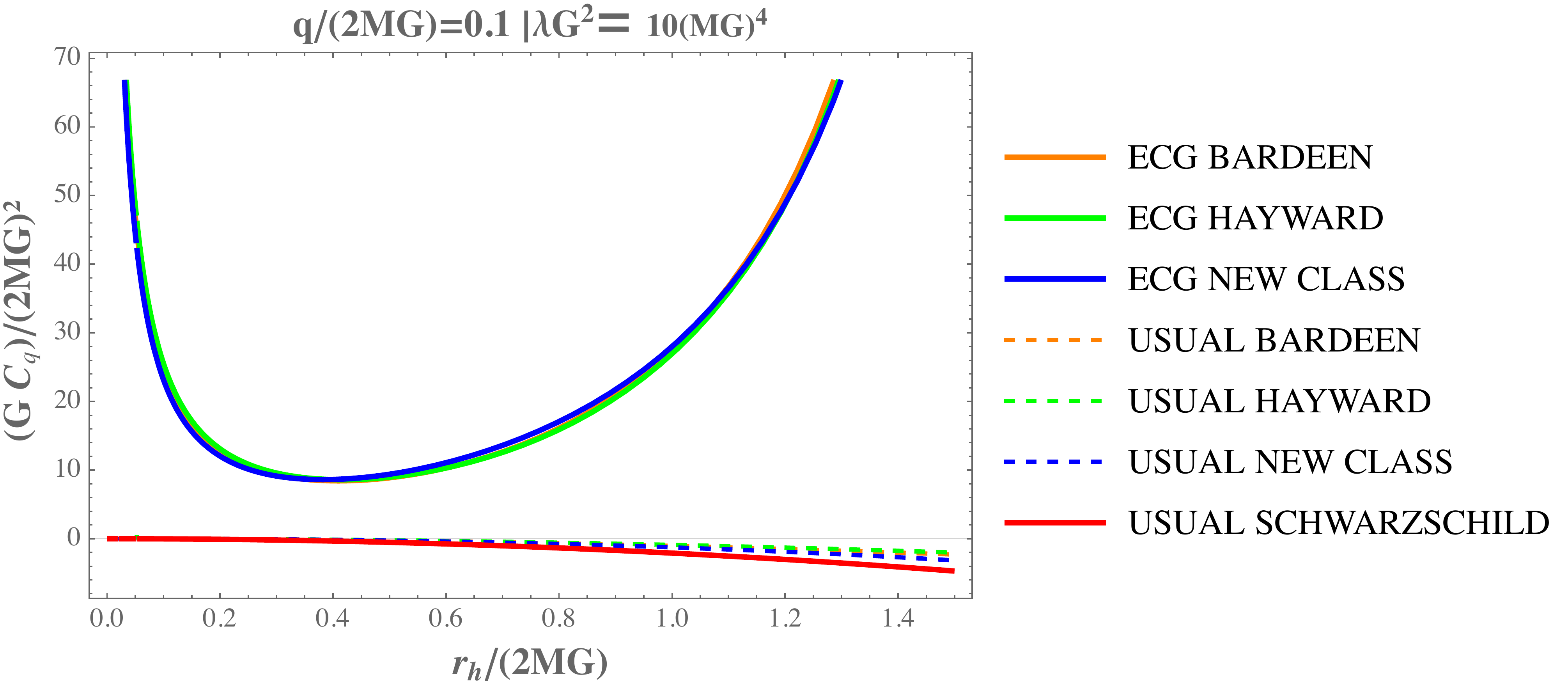}\quad
      \includegraphics[height=4.1cm]{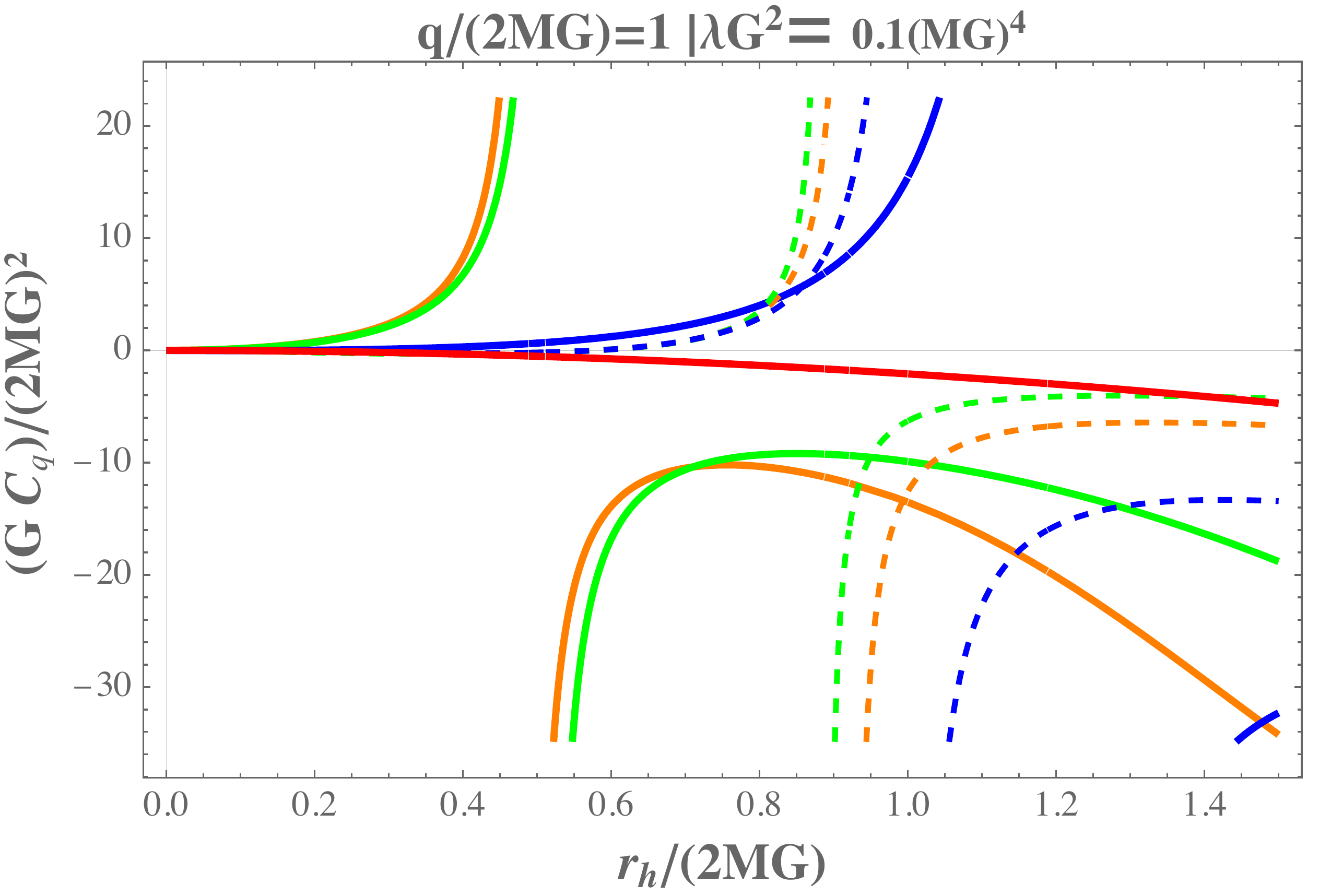}\quad
      \includegraphics[height=4.1cm]{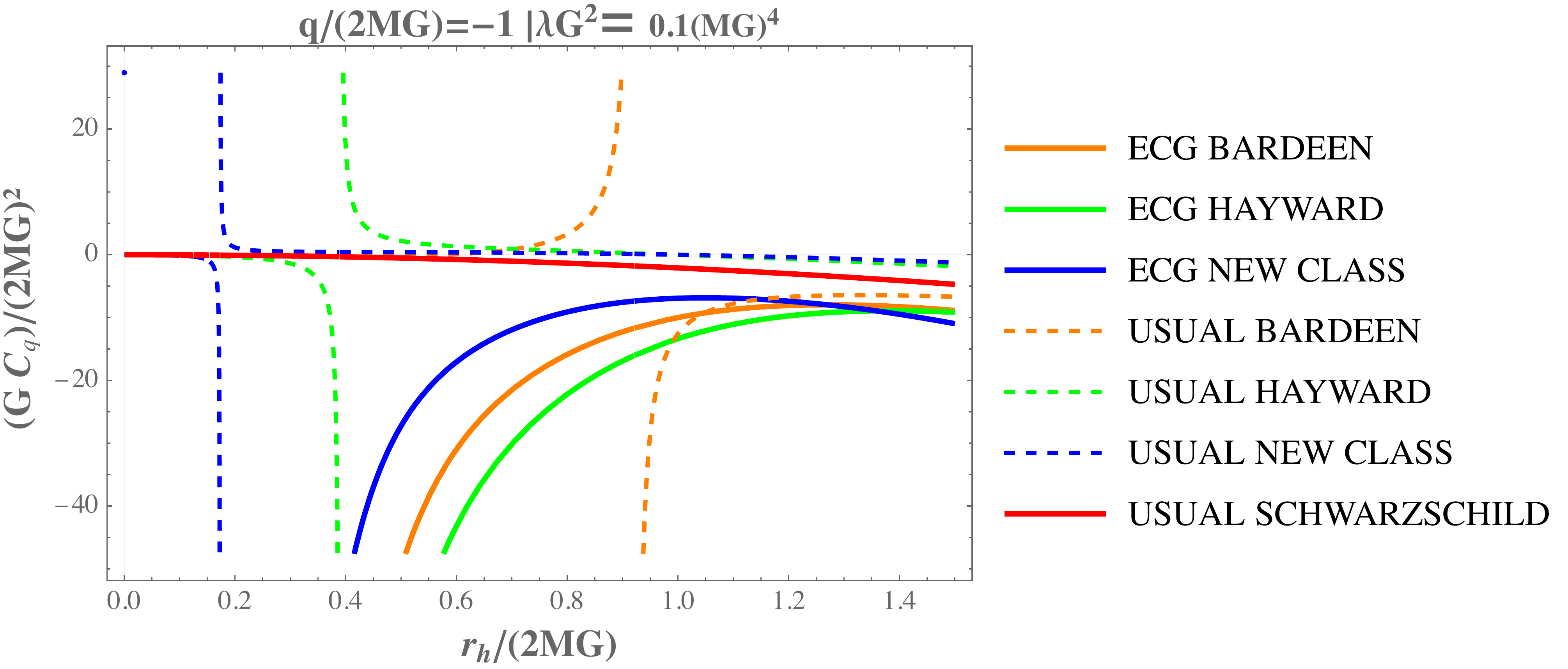}\quad
       \includegraphics[height=4.1cm]{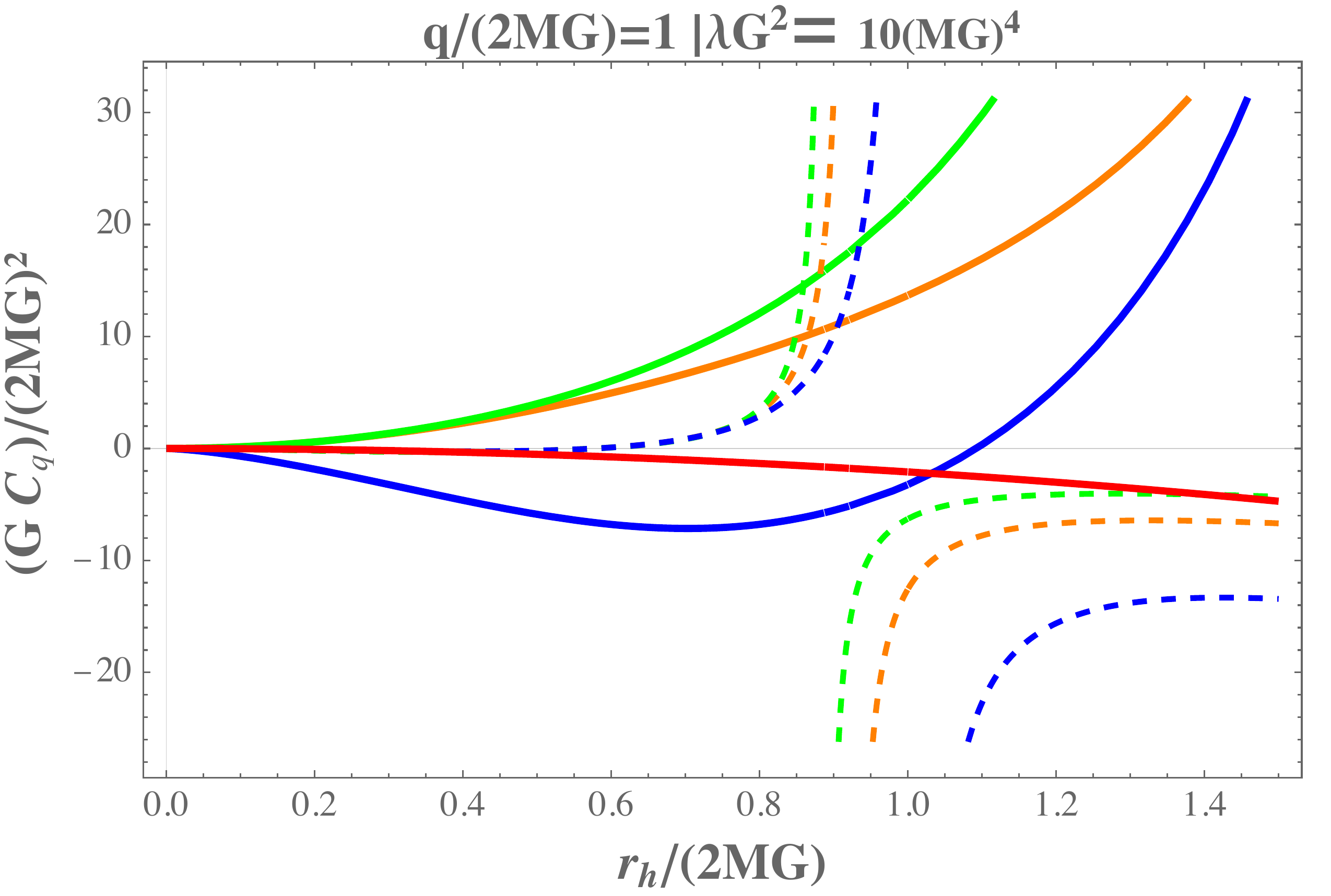}\quad
        \includegraphics[height=4.1cm]{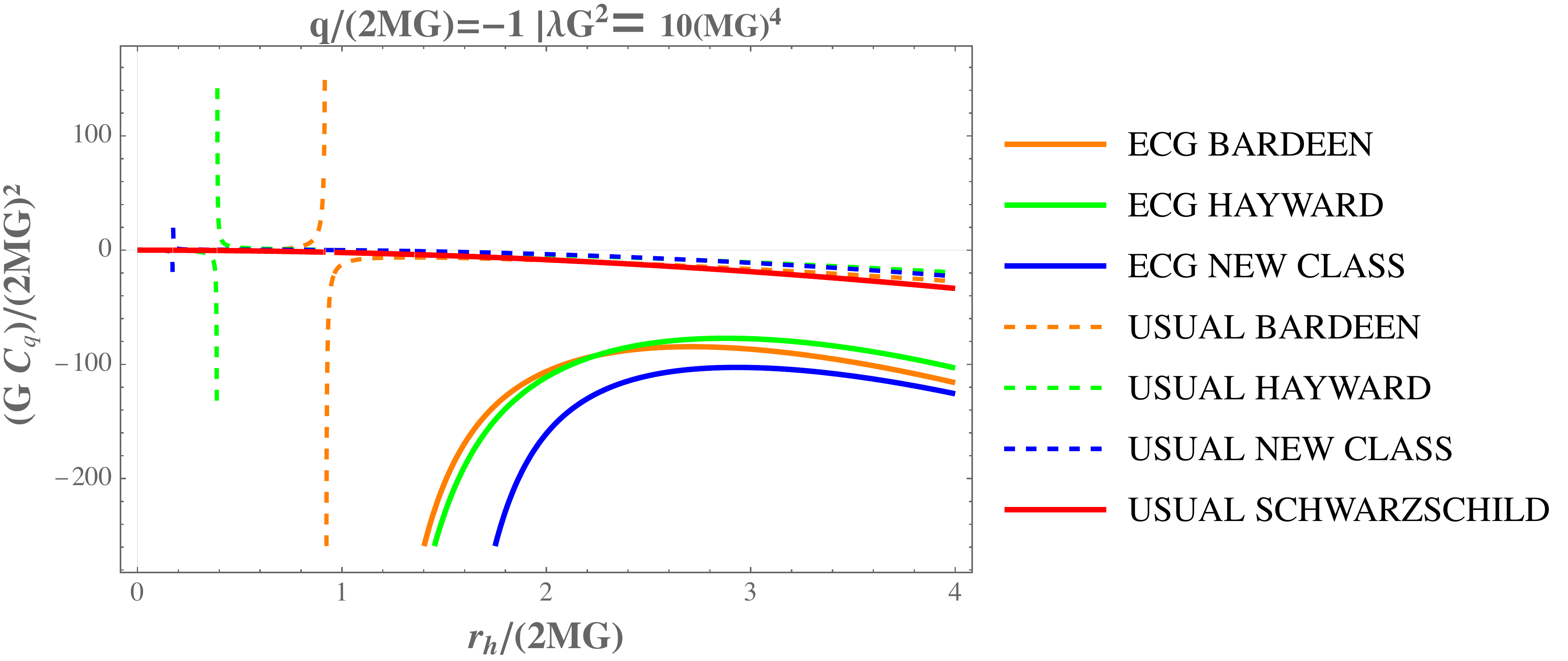}\quad
\caption{The normalized Tempeture $\Tilde{T}= T(2 M G)$  as a function of normalized horizon radius $\Tilde{r}_h$ with small and large $\Tilde{q}$ for various $\lambda$ assuming the Bardeen ($a=3$ and $b=2$), Hayward ($a=b=3$) and New Class ($a=3$ and $b=1$) cases.}  \label{cap}
\end{figure}

After discussed the behaviour of the Hawking temperature, let us now study the thermodynamic stability of the ECG magnetic  solution of Eq.(\ref{fulleom}), by means of the specific heat analysis. Using the Wald entropy obtained in eq.(\ref{entr}) and the Hawking temperature found in eq.(\ref{tem}), we can calculate the specific heat given by
\begin{equation}
    C= T\bigg(\frac{\partial\textbf{S}}{\partial T}\bigg)_q.
\end{equation}
where $q$ is fixed.
Due to the rather cumbersome expression obtained, we employ an graphic analysis of the heat capacity
%We plotted this messy quantity 
for various values of $\lambda$ in Figs[\ref{cap}] assuming that the $q$ can be large or small compared to the Schawzschild radius. 

For $q>0$ (left panel), the fig.\ref{cap} shows a phase transition from a negative into a positive heat capacity for small $q$ and small $\lambda$. Thus, as the black hole emits, it shrinks until a critical horizon radius $r_c$, wherein the positive heat capacity leads to a small stable remnant. Keeping a small magnetic charge and increasing the ECG coupling constant, the black hole undergoes an inverse phase transition leading to a negative heat capacity for small $r_h$. Increasing the ECG coupling, the Bardeen and Hayward solutions recover the negative heat capacity for $r_h \rightarrow\infty$. For a large enough $\lambda$, the three solutions exhibits an exotic divergent positive heat capacity. 

The negative charged solutions (right panel) for small $q$ and $\lambda$ also show a phase transition, though the heat capacity diverges as $r_h \rightarrow 0$. By increasing the charge and the ECG coupling constant, the negative heat capacity for $r_h \rightarrow \infty$ is recovered. However, $C\rightarrow -\infty$ as we approach the origin. As a result, the $q<0$ solutions exhibit instabilities in the both $r_h \rightarrow 0$ and the $r_h \rightarrow \infty$ regimes.

\section{Discussion} 
\label{sec6}

We studied the effects of the Einstein cubic gravity (ECG) on regular black hole solutions. Despite the presence of higher derivatives, the ECG possess the same degrees of freedom of Einstein gravity, at least at the perturbative level. In the strong field regime, previous studies have shown that ECG gravity maintain the singularity at the origin of a static black hole. We added a nonlinear electrodynamics (NLE) lagrangian in order to probe the effects of ECG on the Bardeen, Hayward and new class regular solutions.

For a ECG coupling constant much larger than the magnetic charge of NLE, the regular solutions became singular at the origin. The appearance of naked singularities for even higher ECG constant suggests an upper limit for the ECG in order to satisfy the cosmic censorship conjecture. However, in the perturbative level, the ECG leads to modified regular solutions with the usual De Sitter core. 

We investigated the black hole stability by means of its thermodynamic analysis. The ECG and NLE not only regularize the singularity but also prevents the divergence of the Hawking temperature for small black holes. As shown in the Ref.\cite{pablos}, the temperature vanishes for $r_h \rightarrow 0$ due to the ECG. By adding the NLE source, we found that the temperature vanishes for $r_h\neq 0$ and $M\neq 0$. Thus, the regular black hole leaves a non-emitting and small remnant. The heat capacity revealed that the regular black holes had undergone phases transitions for a critical horizon radius $r_h$.  As the black hole emits and its horizon shrinks, the heat capacity becomes positive and thus, the black hole end up in a final stable remnant. 

The present work suggest further investigations on ECG modified regular black holes. The stability of the perturbed solutions by means of quasinormal modes (QNM) and the phase space are important issues to be addressed. The transition between a regular into a singular NLE black hole due to the non-perturbative ECG is another noteworthy perspective. Moreover, the non trivial form of the entropy points into a general analysis of the first and second law of black hole thermodynamics in this modified gravitational theory, as performed in Ref.\cite{miao} for the $f(T)$ gravity.

%%%%%%%%%%%%%%%%%%%%%%%%%%%%%%%%%%%%%%%%%%%%%%%%%%%%%%%%%%%%%%%%%%%%%%%%%%%%%%%%%%%%%%%%%%%%%%%%%%%%%%%%%%%%%%

\section*{Acknowledgments}
\hspace{0.5cm}The authors thank the Conselho Nacional de Desenvolvimento Cient\'{\i}fico e Tecnol\'{o}gico (CNPq), grants n$\textsuperscript{\underline{\scriptsize o}}$ 304120/2021-9 (JEGS), for financial support.

%%%%%%%%%%%%%%%%%%%%%%%%%%%%%%%%%%%%%%%%%%%%%%%%%%%%%%%%%%%%%%%%%%%%%%%%%%%%%%%%%%%%%%%%%%%%%%%%%%%%%%%%%%%%%%%%%%%%%%%%%%%5

\appendix
\section{ Perturbed equation}
The perturbed equation of (\ref{fulleom}) can be written as
\begin{equation} \label{hap}
    h''(r)+\gamma (r) h'(r)+ \omega^2(r) h(r) =j(r)
\end{equation}
where
\begin{align} \nonumber
  &\gamma (r) = \frac{1}{r \bigg(q^a+r^a\bigg) \bigg((b-3) q^a-3 r^a\bigg) \bigg(r \bigg(q^a+r^a\bigg)^{b/a}-2 G M r^b\bigg)} \bigg[ q^a r^a \bigg(2 G M (a b+b-6) r^b \\  
  &-r((a+5) b-12) \bigg(q^a+r^a\bigg)^{b/a}\bigg)+6 r^{2 a} \bigg(r \bigg(q^a+r^a\bigg)^{b/a}-G M r^b\bigg) +(b-3) q^{2 a} \bigg((b-2) r \bigg(q^a+r^a\bigg)^{b/a}\\ \nonumber 
  &+2 G M r^b\bigg) \bigg], 
\end{align}
\begin{align}\nonumber
    &\omega^2(r) = \frac{1}{24 G^3 \lambda  M \bigg(q^b+r^b\bigg)^2 \bigg((a-3) q^b-3 r^b\bigg) \bigg(2 G M r^a-r \bigg(q^b+r^b\bigg)^{a/b}\bigg)}\bigg[ r^{-a-2} \bigg(-3 q^{2 b} r^b \\ \nonumber
    &\times \bigg(a^2-a (b+8)+9\bigg) r^{a+1} \bigg(q^b+r^b\bigg)^{a/b}+16 G^4 \lambda  M^2 \bigg(a^2 (b+8)-a (5 b+42)+42\bigg) r^{2 a}  \\ \nonumber
    &+r^6 \bigg(q^b+r^b\bigg)^{\frac{2 a}{b}}\bigg)+q^{3 b} \bigg(48 \bigg(a^2-4 a+3\bigg) G^3 \lambda  M r^{a+1} \bigg(q^b+r^b\bigg)^{a/b} +48 \bigg(a^3-8 a^2\bigg) G^4 \lambda  M^2 r^{2 a}  \\ \nonumber
    & +48 \bigg(21 a-14\bigg) G^4 \lambda  M^2 r^{2 a} -r^6 \bigg(q^b+r^b\bigg)^{\frac{2 a}{b}}\bigg) -3 q^b r^{2 b} \bigg(16 G^3 \lambda  M (a (b+4)-9) r^{a+1} \bigg(q^b+r^b\bigg)^{a/b}\\ \nonumber
    & +r^6 \bigg(q^b+r^b\bigg)^{\frac{2 a}{b}} -16 G^4 \lambda  M^2 (a (5 b+21)-42) r^{2 a}\bigg) -r^{3 b} \bigg(-144 G^3 \lambda  M r^{a+1} \bigg(q^b+r^b\bigg)^{a/b} \\ \nonumber
    &+r^6 \bigg(q^b+r^b\bigg)^{\frac{2 a}{b}}+672 G^4 \lambda  M^2 r^{2 a}\bigg)\bigg) \bigg]
\end{align}
and
\begin{align} \nonumber
    &j(r)= - \frac{1}{72 G^3 \lambda  M \bigg((a-3) q^b-3 r^b\bigg) \bigg(2 G M r^a-r \bigg(q^b+r^b\bigg)^{a/b}\bigg)}\bigg[  r^{-a-3} \bigg(q^b+r^b\bigg)^{-\frac{a}{b}-2} \\ \nonumber
    &\times\bigg(-3 q^{2 b} r^b \bigg(96 G^5 \lambda  M^3 \bigg(a^2 (b+5)-3 a (b+8)+23\bigg) r^{3 a}-48 (a-3) G^4 \lambda  M^2 (a (b+7)-9) \\ \nonumber
    &\times r^{2 a+1}\bigg(q^b+r^b\bigg)^{a/b}+q^{3 b} \bigg(96 \bigg(2 a^3-15 a^2+36 a-23\bigg) G^5 \lambda  M^3 r^{3 a}-144 (a-3)^2 (a-1) G^4 \lambda  M^2 \\ \nonumber
    &\times r^{2 a+1} \bigg(q^b+r^b\bigg)^{a/b}+3 q^b r^{2 b} \bigg(-144 G^4 \lambda  M^2 (a (b+5)-9) r^{2 a+1} \bigg(q^b+r^b\bigg)^{a/b} \\ \nonumber
    &+96 G^5 \lambda  M^3 (3 a (b+4)-23) r^{3 a}\bigg)+r^{3 b}\bigg(1296 G^4 \lambda  M^2 r^{2 a+1} \bigg(q^b+r^b\bigg)^{a/b}-2208 G^5 \lambda  M^3 r^{3 a}\bigg)\bigg)\bigg]
\end{align}
Note that the limit case $q=0$, we recover the homogeneous equation obtain by Ref. [\cite{pablos}]. As already mentioned in the text, it is easy to show that for the limit of large r (asymptotic limit) the coefficients $\gamma (r)$ and $\omega^2(r)$ coincide with those obtained in Ref[\cite{pablos}]. However, in the limit of small r (near the origin), it is necessary to specify which nonlinear electrodynamics we are dealing with. For the Bardeen case, we obtain that the leading terms in the origin are given by Eq. (\ref{g2}) and (\ref{w2}). On the other hand, the Hayward case with $a=b=3$, the coefficients are given by (\ref{g3}) and (\ref{w3}). Finally, in the case of $a=3$ and $b=1$, they are given by (\ref{g4}) and (\ref{w4}).
\section{Black hole entropy}
We know that the theory of high derivatives of gravity normally does not obey Bekenstein–Hawking area law for calculating entropy. In this case we need to use Wald's formula \cite{w1,w2}. Thus, the entropy associated with a static and symmetrically spherical black hole in ECG is given by
 \begin{equation}\label{for}
     \textbf{S} = -2\pi \int _{\Sigma} d ^2x \sqrt{h}\frac{\delta \mathcal{L} }{\delta R_{\mu\nu\alpha\beta}}\epsilon_{\mu\nu}\epsilon_{\alpha\beta},
 \end{equation}
where $\frac{\delta  }{\delta R_{\mu\nu\alpha\beta}}$ is the Euler-Lagrange derivative, $h$ is the determinat of the induced metric on the horizon $\Sigma$ and $\epsilon_{\mu\nu}$ is the binormal of the horizon. 

Applying the formula (\ref{for}) to action (\ref{action})  and assuming ansatz (\ref{metric}) ($N'(r)=0$), we find that
 \begin{equation}\label{for}
     \textbf{S} = \frac{\pi r_h ^2}{G}\bigg[1 - \frac{48G^2\lambda\kappa_g^2}{r_h ^2}\bigg( \frac{2}{\kappa_g r_h} + 1\bigg) \bigg]
 \end{equation}
where we use the fact that $f'(r_h)=2\kappa_g$. Substituting Eq. (\ref{sg}) in the expression above, we find that the entropy is given by
 \begin{align} \label{entr} 
    \textbf{S} = \frac{\pi  r_h^2}{G} -\frac{48 \pi  G \lambda  \left(r_h- \frac{2 a G M r_h^a q^b}{\left(q^b+r_h^b\right)^{\frac{a+b}{b}}} \right)^2}{r_h ^4 \left(\sqrt{\frac{48 G^2 \lambda  \left(1- \frac{2 a G M r_h^{a-1} q^b}{\left(q^b+r_h^b\right)^{\frac{a+b}{b}}} \right)}{r_h^4}+1}+1\right)^2}  \left(\frac{2 r_h \left(\sqrt{\frac{48 G^2 \lambda  \left(1-\frac{2 a G M r_h^{a-1} q^b}{\left(q^b+r_h^b\right)^{\frac{a+b}{b}}} \right)}{r_h^4}+1}+1\right)}{r_h-\frac{2 a G M r_h^a q^b}{\left(q^b+r_h^b\right)^{\frac{a+b}{b}}} }+1\right)
 \end{align}
\end{document}